\documentclass[journal,final]{IEEEtran}
\usepackage{amsmath,amssymb,amsfonts}
\usepackage{graphicx}
\usepackage{subfigure}
\usepackage{booktabs}
\usepackage{array}
\usepackage{cite}
\usepackage[ruled]{algorithm2e}

\begin{document}

\title{Diagnosis of Intelligent Reflecting Surface in Millimeter-wave Communication Systems}

\author{
Rui Sun,
Weidong Wang,
Li Chen,
Guo Wei, and
Wenyi Zhang, \IEEEmembership{Senior Member, IEEE}
	
\thanks{
The authors are with the CAS Key Laboratory of Wireless-Optical Communications, University of Science and Technology of China, Hefei 230027, China (e-mail: ruisun@mail.ustc.edu.cn; wdwang@ustc.edu.cn; chenli87@ustc.edu.cn; wei@ustc.edu.cn; wenyizha@ustc.edu.cn).

Codes available on: https://github.com/DestinationSR/IRSdiagnosis.
}
}

\maketitle

\begin{abstract}
Intelligent reflecting surface (IRS) is a promising technology for enhancing wireless communication systems.
It adaptively configures massive passive reflecting elements to control wireless channel in a desirable way.
Due to hardware characteristics and deploying environments, an IRS may be subject to reflecting element blockages and failures, and hence developing diagnostic techniques is of great significance to system monitoring and maintenance.
In this paper, we develop diagnostic techniques for IRS systems to locate faulty reflecting elements and retrieve failure parameters.
Three cases of channel state information (CSI) availability are considered.
In the first case where full CSI is available, a compressed sensing based diagnostic technique is proposed, which significantly reduces the required number of measurements.
In the second case where only partial CSI is available, we jointly exploit the sparsity of the millimeter-wave channel and the failure, and adopt compressed sparse and low-rank matrix recovery algorithm to decouple channel and failure.
In the third case where no CSI is available, a novel atomic norm is introduced as the sparsity-inducing norm of the cascaded channel, and the diagnosis problem is formulated as a joint sparse recovery problem.
Finally, the proposed diagnostic techniques are validated through numerical simulations.
\end{abstract}


\section{Introduction}
Intelligent reflecting surface (IRS) is an emerging technology to improve the performance of wireless communication systems.
By adaptively adjusting the attenuation and/or phase shift of massive passive reflecting elements, the IRS is able to ``program" the wireless propagation channel with low-cost hardware, which shows great potential for improving spectrum efficiency and energy efficiency, especially for millimeter-wave (mmWave) communication system due to its vulnerability to high path-loss and link blockage \cite{wu2019towards, di2020smart}.
Massive research works on the IRS have been conducted recently, including channel modeling \cite{tang2020wireless, di2020analytical}, beamforming design \cite{wu2019beamforming, zhao2020intelligent}, and hardware experiments \cite{tang2020mimo, dai2020reconfigurable}.

In the meantime, hardware characteristics and deploying environments also bring unique reliability challenges to the IRS.
The IRS consists of massive passive ``meta-atoms" that are made of two-dimensional (2D) metamaterial \cite{cui2014coding, liaskos2018new}, which may result in a high failure probability due to large system scale and dense interconnections of tuning circuits \cite{kouzapas2020towards, taghvaee2020error}.
On the other hand, the short wavelength of mmWave band makes the IRS vulnerable to the blockages with comparable sizes like dirt, particles, and flying debris in outdoor deploying environments \cite{eltayeb2018compressive}.
Reflecting characteristics of the IRS also make it more sensitive to failures than conventional MIMO systems.
It is reported in \cite{liang2019large} that the power gain provided by the IRS scales up with the square of the number of reflecting elements.
Therefore, the IRS also suffers from ``squared power loss" since faulty reflecting elements could cause twice attenuation during incidence and reflection.
It is also reported in \cite{taghvaee2020error} that faulty reflecting elements may result in various radiation pattern distortion issues like degraded directivity, increased side-lobe level, and beam misalignment.
In summary, the existence of faulty reflecting elements could significantly degrade system performance, which necessitates diagnostic techniques for locating faulty reflecting elements and retrieving failure parameters for calibration purposes.

To provide references for the IRS diagnosis, we refer to extensive related works on antenna array diagnosis.
Early representative works include \cite{lee1988near} and \cite{bucci2005accurate}.
These techniques exploit the linear relationship between antenna pattern and array excitation, and take samples uniformly on the radiation pattern to recover actual array excitation by solving an inverse problem.
Therefore, faulty antenna elements can be identified by comparing the actual array excitation and the ideal one.
In order to make the inverse problem solvable, these techniques require the number of measurements no less than the total number of antenna elements, which may be time-consuming for large-scale systems.
About a decade ago, compressed sensing was introduced to array diagnosis \cite{migliore2011compressed}, which efficiently captures the failure sparsity that the number of faulty antenna elements is usually much smaller than the total number of antenna elements.
Hence, one can subtract the ideal antenna pattern from the observed one to obtain the differential pattern.
The resulting differential array excitation is a sparse signal that can be recovered from compressed measurements, in which non-zero entries indicate faulty antenna elements.
That work has initiated numerous compressed sensing based diagnostic techniques.
One type of improvement focuses on the enhancement in sparse recovery algorithms, such as Bayesian compressive sensing \cite{oliveri2012reliable, salucci2018planar}, non-convex compressed sensing \cite{ince2015array}, and mixed-norm constrained compressed sensing \cite{fuchs2016fast}.
Other improvements are mainly the development of new sampling methods, including far-field sampling \cite{migliore2013array}, phase-less sampling \cite{palmeri2019diagnosis}, phase domain sampling \cite{xiong2019compressed}, and random beam sampling \cite{eltayeb2018compressive, aldayel2019constrained, ma2020antenna}.
In particular, the work \cite{sun2021hybrid} extended the above antenna-only diagnoses to the joint diagnosis of antenna, phase shifter, and RF chain of hybrid beamforming (HBF) system.

The above pioneering works provide various approaches to detect faulty antenna elements in conventional MIMO systems.
However, they do not directly apply to the IRS diagnosis since they fail to address the following challenges.
The first challenge is the coupling between channel and failure.
Intuitively speaking, it is difficult to distinguish antenna failure from channel fading since an antenna failure can be treated as a part of channel.
To cope with this issue, almost all of currently proposed diagnostic techniques assume a free-space wireless environment in order to calculate the channel using known spatial coordinates, which is then used to generate fault-free reference signal for detecting failures.
Such requirement significantly limits their applications in outdoor online diagnosis since wireless channel is subject to inevitable multipath scattering, which makes accurate reference signal unavailable.
The second challenge arises from the hardware characteristic of the IRS.
The IRS is a passive device that has no signal processing capability except for configurable reflecting coefficients.
Hence, one has no direct access to the received signal of each reflecting element.
To perform diagnosis, auxiliary transmitter (TX) and receiver (RX) are required to play the role of ``signal generator" and ``signal analyzer", respectively.
The equivalent channel from the TX through the IRS to the RX is a cascaded channel, which combines the effect of multipath scattering in both TX-IRS and IRS-RX channels.
Such channel characteristic makes the channel-failure coupling issue even worse and introduces more channel uncertainties to the diagnosis.
Current literature fails to address the above issues.
This motivates us to develop tailored diagnostic techniques that incorporate the hardware and channel characteristics of an IRS system.

In this paper, we develop diagnostic techniques for an IRS system to locate faulty reflecting elements and retrieve failure parameters.
We first establish the IRS failure model and discuss the channel state information (CSI) availability.
Specifically, three cases of the CSI availability are considered and three different diagnostic techniques are developed accordingly.
For the diagnosis with full CSI (all channels are known), a compressed sensing based diagnostic technique is proposed, which significantly shortens the diagnostic time by exploiting the failure sparsity.
For the diagnosis with partial CSI (only the channel from the IRS to the RX is known), the channel-failure coupling issue is solved by jointly exploiting the sparsities of failure and mmWave channel, and the diagnosis problem is formulated as a compressed sparse and low-rank matrix recovery (cSLRMR) problem.
For the diagnosis without CSI (both channels are unknown), a novel atomic norm is introduced to induce the sparsity of cascaded channel, and an atomic norm minimization (ANM) based diagnostic technique is proposed to decouple the cascaded channel and failure.
The proposed ANM-based diagnostic technique is further extended to the multiple measurement vectors (MMV) model to exploit the diversity gain provided by multipath scattering.
Two efficient algorithms are also proposed to accelerate the diagnosis.
The main contributions of this paper can be summarized as follows:
\begin{itemize}
\item \textbf{IRS failure modeling.}
We establish the failure model for an IRS system, which incorporates the attenuation and phase-shifting effects of reflecting element failure.
The proposed failure model is further extended to the multi-antenna setting to exploit the diversity gain provided by multipath scattering.

\item \textbf{Channel-failure decoupling.} 
The coupling between channel fading and failure is a major challenge of IRS diagnosis.
To overcome this issue, we jointly exploit the sparsity of mmWave channel and failure, and propose a cSLRMR-based diagnostic technique and an ANM-based diagnostic technique for the case of partial CSI and no CSI, respectively.

\item \textbf{Efficient diagnostic algorithms.}
The proposed ANM-based diagnostic technique has strong robustness against multipath scattering, but it also suffers from high computational complexity if conventional solvers are adopted.
To accelerate the diagnosis, we develop alternating direction method of multipliers (ADMM) based algorithms to implement the proposed techniques, which significantly shorten the diagnostic time compared with conventional algorithms.
\end{itemize}

The remaining part of this paper is organized as follows.
Section \ref{sec:sysModel} describes system failure model and channel model.
Section \ref{sec:caseIfCSI}, Section \ref{sec:caseIIpCSI}, and Section \ref{sec:caseIIInCSI} present the diagnostic techniques that require full CSI, partial CSI, and no CSI, respectively.
Section \ref{sec:simu} presents some illustrative instances and numerical results.
Finally, Section \ref{sec:conclu} concludes the paper.

\begin{table}[t]
	\small
	\centering
	\caption{Notations adopted in this paper.}
	\label{nota}
	\begin{tabular}{p{1.5cm}<{\raggedright} p{6cm}<{\raggedright}}
		\toprule
		\textbf{Notation} & \textbf{Description} \\
		\midrule
		$a,A$ & a scalar \\
		$\mathbf{a}$ & a vector \\
		$\mathbf{A}$ & a matrix \\
		$[\mathbf{a}]_i$ & $i$-th element of vector $\mathbf{a}$ \\
		$[\mathbf{A}]_{i,j}$ & the element in $i$-th row and $j$-th column of matrix $\mathbf{A}$ \\
		$[\mathbf{A}]_{i,:}$ & $i$-th row of $\mathbf{A}$ \\
		$\mathbf{I}_N$ & an $N \times N$ identity matrix \\
		$\mathbf{1}_{M \times N}$ & an all-one matrix of size $M \times N$ \\
		$\Vert \mathbf{a} \Vert_p$ & $l_p$ norm of $\mathbf{a}$ \\
		$\Vert \mathbf{a} \Vert_\mathcal{A}$ & atomic norm of $\mathbf{a}$ \\
		$\Vert \mathbf{A} \Vert_\mathrm{F}$ & Frobenius norm of $\mathbf{A}$ \\
		$\Vert \mathbf{A} \Vert_*$ & nuclear norm of $\mathbf{A}$ (sum of the singular values of $\mathbf{A}$) \\
		$(\cdot)^\mathrm{T}$ & transpose \\
		$(\cdot)^\mathrm{H}$ & conjugate transpose \\
		$\mathrm{vec}(\cdot)$ & vectorization \\
		$\mathrm{diag}(\mathbf{a})$ & a diagonal matrix with the elements of $\mathbf{a}$ on its main diagonal \\
		$\mathrm{tr}(\cdot)$ & matrix trace \\
		$\circ$ & Hadamard product (element-wise product) \\
		$\oslash$ & Hadamard division (element-wise division) \\
		$\otimes$ & Kronecker product \\
		$\mathcal{CN}(\mu, \sigma^2)$ & circularly symmetric complex Gaussian distribution with mean $\mu$ and variance $\sigma^2$ \\
		$U(a,b)$ & uniform distribution over the interval $(a,b)$ \\
		\bottomrule
	\end{tabular}
\end{table}

The notations adopted in this paper are listed in Table \ref{nota}.

\section{System Model}
\label{sec:sysModel}
In this section, we present system failure model and discuss the CSI availability of three cases.

\subsection{Failure Model}
\begin{figure}
	\centering
	\includegraphics[scale=0.45]{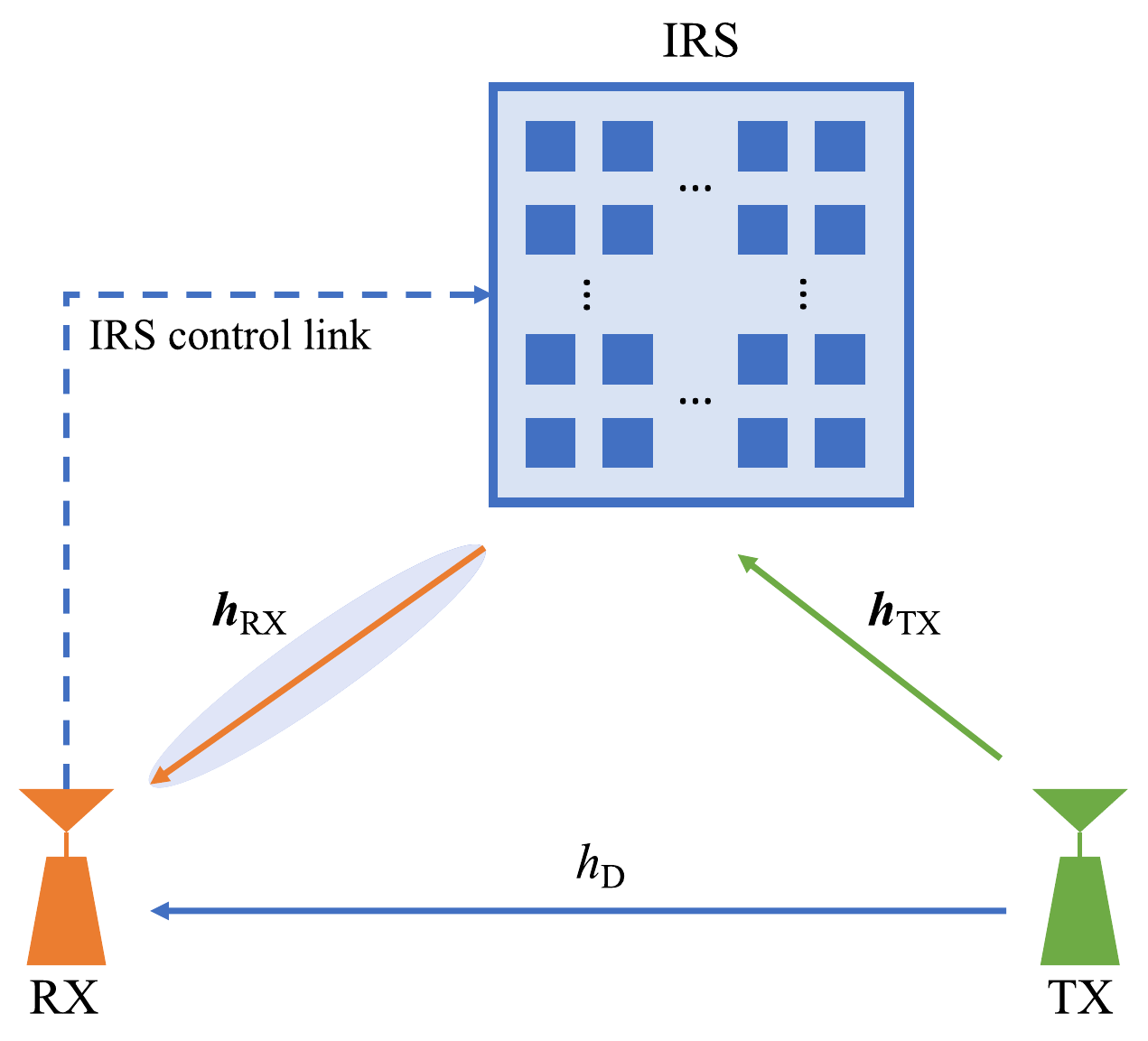}
	\caption{Diagnosis setup.}
	\label{systemModel}
\end{figure}

\begin{figure}
	\centering
	\includegraphics[scale=0.45]{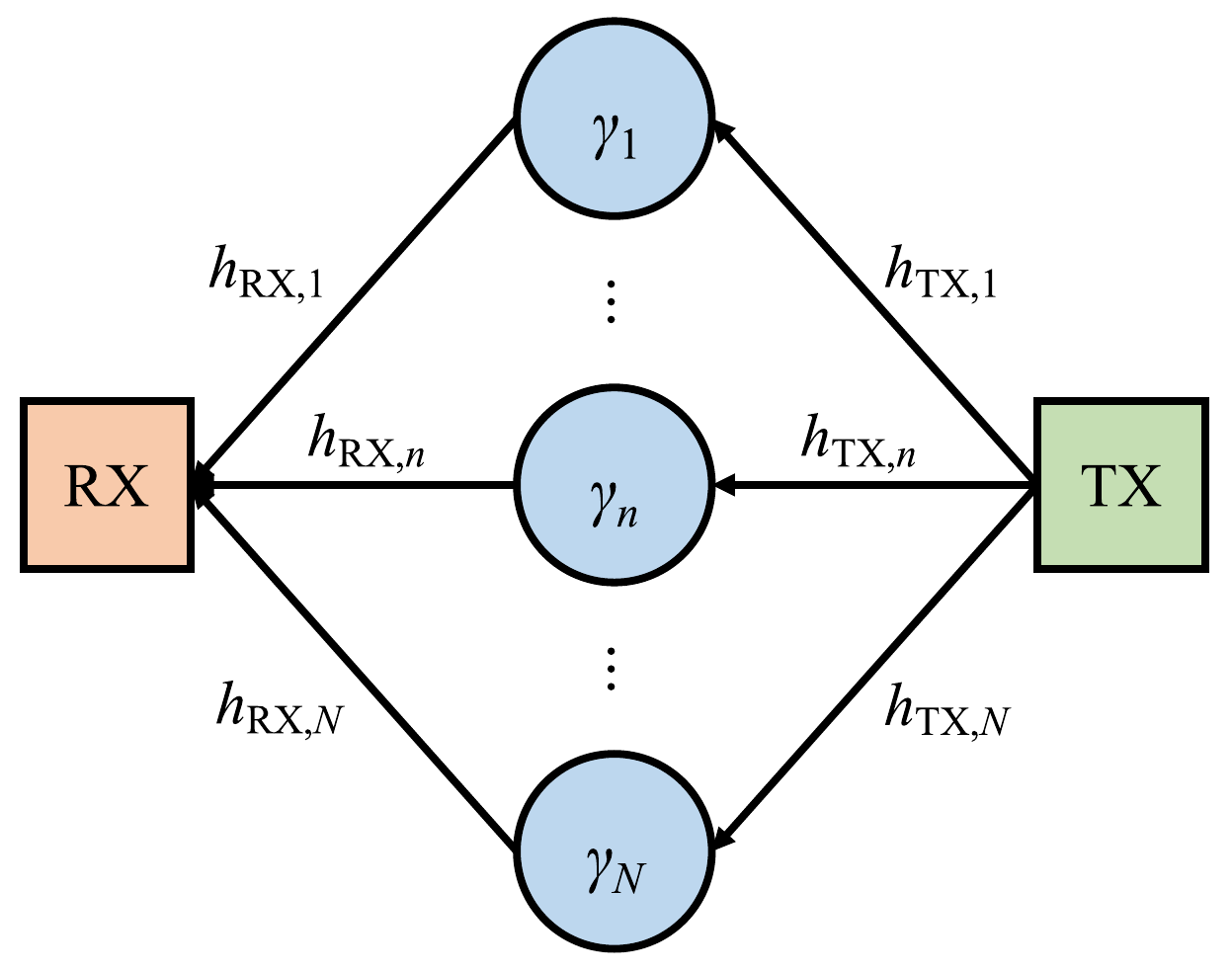}
	\caption{IRS weighting scheme, where $h_{\mathrm{TX},n}$ and $h_{\mathrm{RX},n}$ are the $n$-th channel coefficient in $\mathbf{h}_\mathrm{TX}$ and $\mathbf{h}_\mathrm{RX}$, respectively.}
	\label{nodeFig}
\end{figure}

The diagnosis setup is shown in Fig. \ref{systemModel}, where the IRS is operating at the mmWave band and is the system to be diagnosed.
The IRS is in uniform planar array (UPA) structure and consists of $N = H \times W$ reflecting elements, where $H$ and $W$ are the number of elements along vertical and horizontal directions, respectively.
Both the TX and the RX are equipped with a single antenna.\footnote{The extension to multi-antenna RX will be presented in Sec. \ref{sec:multiRX}.}
The fault-free received signal at the RX can be expressed as \cite{wu2019intelligent}
\begin{equation}
\label{tansModel}
y' = \left(\mathbf{h}_\mathrm{RX}^\mathrm{T} \boldsymbol{\Gamma} \mathbf{h}_\mathrm{TX} + h_\mathrm{D} \right) x + w',
\end{equation}
where $\mathbf{h}_\mathrm{RX} \in \mathbb{C}^N$ and $\mathbf{h}_\mathrm{TX} \in \mathbb{C}^N$ are the channel vectors from the IRS to the RX and from the TX to the IRS, respectively;
$h_\mathrm{D} \in \mathbb{C}$ is the direct channel coefficient from the TX to the RX;
$x$ and $w'$ are the transmit symbol and the noise, respectively;
$\boldsymbol{\Gamma} \in \mathbb{C}^{N \times N} = \mathrm{diag}(\mathbf{f}) = \mathrm{diag}([\beta_1 e^{j \gamma_1}, \cdots, \beta_N e^{j \gamma_N}]^\mathrm{T})$ is the weighting matrix of the IRS, where $\beta_n$ and $\gamma_n$ are the attenuation and phase shift of the $n$-th reflecting element, respectively.
To maximize reflection power, we only adjust phase shift and set $\beta_n = 1$ for $n=1, \cdots, N$ throughout the diagnosis.
An illustrative figure of the IRS weighting scheme is shown in Fig. \ref{nodeFig}.

To represent faulty reflecting elements, we multiply each reflecting coefficient by a failure mask $m_n \in \mathbb{C}, n=1, \cdots N$ \cite{eltayeb2018compressive, taghvaee2020error}, which is defined as
\begin{equation}
	m_n =
	\begin{cases}
		1, & n\text{-th reflecting element is functioning} \\
		\eta_n e^{j\kappa_n}, & n\text{-th reflecting element is faulty}
	\end{cases},
\end{equation}
where $0 \le \eta_n < 1$ and $0 \le \kappa_n < 2\pi$ represent the failure induced attenuation and phase shift of the $n$-th reflecting element, respectively.
Therefore, the transmission model \eqref{tansModel} can be rewritten as
\begin{equation}
	\label{tansModel2}
	y' = \left(\mathbf{h}_\mathrm{RX}^\mathrm{T} \bigg(\boldsymbol{\Gamma} \mathrm{diag}(\mathbf{m})\bigg) \mathbf{h}_\mathrm{TX} + h_\mathrm{D} \right) x + w',
\end{equation}
where $\mathbf{m}= [m_1, m_2, \cdots, m_N]^\mathrm{T}$ is the vector of failure masks.
We assume that a \textit{failure sparsity} property holds; that is, the number of faulty reflecting elements is much smaller than the total number of elements \cite{migliore2011compressed, eltayeb2018compressive, xiong2019compressed}, and hence most entries in $\mathbf{m}$ are unity.\footnote{
In practice, the failure masks $\mathbf{m}$ may not be a near all-one vector due to general hardware imperfections (i.e., $\mathbf{m}-\mathbf{1}_{N \times 1}$ may not be strictly sparse).
In view of that, we focus on detecting faults and treat hardware imperfections as additive noises in the measurement or Gaussian perturbations on the measurement matrix.
As analyzed in \cite{parker2011compressive}, a simple and effective way to cope with this issue is to adjust the regularization parameters in the proposed diagnostic techniques to handle the increased noise level.}

For the purpose of diagnosis, the TX is set to transmit symbol $x=1$ throughout the diagnosis.
Besides, to eliminate the received signal contributed by the direct channel $h_\mathrm{D}$, we first set the IRS to the absorbing mode (i.e., $\beta_n = 0$ for $n=1, \cdots, N$) before the diagnosis and conduct channel estimation to obtain an estimator of $h_\mathrm{D}$ as $\hat{h}_\mathrm{D}$ \cite{wu2019towards}, and then subtract it from $y'$.
Therefore, the transmission model for diagnosis can be rewritten as
\begin{equation}
\label{sysModelAll}
\begin{aligned}
y 
&= y'|_{x=1} - \hat{h}_\mathrm{D} \\
&= \mathbf{h}_\mathrm{RX}^\mathrm{T} \bigg(\boldsymbol{\Gamma} \mathrm{diag}(\mathbf{m})\bigg) \mathbf{h}_\mathrm{TX} + \underbrace{h_\mathrm{D} - \hat{h}_\mathrm{D} + w'}_{=w} \\
&= \mathbf{h}_\mathrm{RX}^\mathrm{T} \bigg(\mathrm{diag}(\mathbf{f}) \mathrm{diag}(\mathbf{m}) \bigg) \mathbf{h}_\mathrm{TX} + w \\
&= \sum_{n=1}^{N} \left[\mathbf{h}_\mathrm{RX} \circ \mathbf{f} \circ \mathbf{m} \circ \mathbf{h}_\mathrm{TX}\right]_n + w,
\end{aligned}
\end{equation}
where $w \sim \mathcal{CN}(0, 1/\mathrm{SNR})$ is the augmented noise due to the noisy estimation of $h_\mathrm{D}$ and $\mathrm{SNR}$ is the signal-to-noise ratio.

From received signal $y$, the goal of diagnosis is to recover the failure mask vector $\mathbf{m}$, in which non-unity entries indicate faulty reflecting elements.
Besides, recovered failure masks can be utilized for calibration, which has been treated in \cite{mailloux1996array, yeo1999array, li2009source} and references therein.

In \eqref{sysModelAll}, $\mathbf{f}$ is the configurable weighing vector and $\mathbf{m}$ is the failure mask to be estimated, while $\mathbf{h}_\mathrm{RX}$ and $\mathbf{h}_\mathrm{TX}$ are unknown channel vectors that need to be estimated based on field measurements.
Unfortunately, conventional channel estimation techniques like channel training become invalid since channel vectors and failure masks are coupled (as shown in \eqref{sysModelAll}).
At first glance, it seems impossible to distinguish channel fading from reflecting element failure, and hence the CSI availability brings a major challenge to IRS diagnosis.
In the following, we will discuss how to deal with unknown CSI in three cases.

\subsection{Channel Model}
\label{sec:chanModel}
To better understand our channel modeling assumptions, we first introduce mmWave narrowband clustered channel model \cite{el2014spatially}
\begin{equation}
\begin{aligned}
	\label{chanModel}
	\mathbf{h}_\mathrm{TX}
	&= \mathrm{vec}(\mathbf{H}_\mathrm{TX}) \\
	&= \mathrm{vec} \left( \sum_{l=1}^{L_\mathrm{TX}} \alpha_{\mathrm{TX},l} \mathbf{a}_\mathrm{H}(\theta_{\mathrm{TX},l},\phi_{\mathrm{TX},l}) \mathbf{a}_\mathrm{W}^\mathrm{T}(\theta_{\mathrm{TX},l},\phi_{\mathrm{TX},l}) \right) \\
	&= \sum_{l=1}^{L_\mathrm{TX}} \alpha_{\mathrm{TX},l} \bigg( \mathbf{a}_\mathrm{W}(\theta_{\mathrm{TX},l},\phi_{\mathrm{TX},l}) \otimes \mathbf{a}_\mathrm{H}(\theta_{\mathrm{TX},l},\phi_{\mathrm{TX},l}) \bigg), \\
	\mathbf{h}_\mathrm{RX}
	&= \mathrm{vec}(\mathbf{H}_\mathrm{RX}) \\
	&= \mathrm{vec} \left( \sum_{l=1}^{L_\mathrm{RX}} \alpha_{\mathrm{RX},l} \mathbf{a}_\mathrm{H}(\theta_{\mathrm{RX},l},\phi_{\mathrm{RX},l}) \mathbf{a}_\mathrm{W}^\mathrm{T}(\theta_{\mathrm{RX},l},\phi_{\mathrm{RX},l}) \right) \\
	&= \sum_{l=1}^{L_\mathrm{RX}} \alpha_{\mathrm{RX},l} \bigg( \mathbf{a}_\mathrm{W}(\theta_{\mathrm{RX},l},\phi_{\mathrm{RX},l}) \otimes \mathbf{a}_\mathrm{H}(\theta_{\mathrm{RX},l},\phi_{\mathrm{RX},l}) \bigg),
\end{aligned}
\end{equation}
where 
\begin{equation}
\begin{aligned}
\mathbf{a}_\mathrm{H}(\theta,\phi) &= [1, e^{j2\pi d \sin\theta \sin \phi}, \cdots, e^{j2\pi d (H-1) \sin\theta \sin \phi}]^\mathrm{T}, \\
\mathbf{a}_\mathrm{W}(\theta,\phi) &= [1, e^{j2\pi d \sin\theta \cos \phi}, \cdots, e^{j2\pi d (W-1) \sin\theta \cos \phi}]^\mathrm{T},
\end{aligned}
\end{equation}
and $d=1/2$ is the element spacing normalized by the wavelength;
$L_\mathrm{TX}$ is the number of sub-paths in the TX-IRS channel;
$\alpha_{\mathrm{TX},l}$, $\theta_{\mathrm{TX},l}$, and $\phi_{\mathrm{TX},l}$ are the complex gain, the elevation angle of arrival (AOA), and the azimuth AoA of the $l$-th sub-path in the TX-IRS channel, respectively.
The parameters associated with the IRS-RX channel are defined similarly using the subscript ``RX'' and the term ``angle of departure (AoD)".
In addition, we assume a block fading channel that the channel remains constant within a coherence interval \cite{el2014spatially}.

We consider the above channel model in three cases: full CSI (both $\mathbf{h}_\mathrm{RX}$ and $\mathbf{h}_\mathrm{TX}$ are available), partial CSI (only $\mathbf{h}_\mathrm{RX}$ is available), and no CSI (both $\mathbf{h}_\mathrm{RX}$ and $\mathbf{h}_\mathrm{TX}$ are unavailable).

\subsubsection{Case I (Full CSI)}
For factory test, the diagnosis is usually performed in a microwave anechoic chamber where the wireless environment can be precisely controlled.
Therefore, the channel vectors $\mathbf{h}_\mathrm{RX}$ and $\mathbf{h}_\mathrm{TX}$ can be calculated based on known spatial coordinates \cite{migliore2013array, fuchs2016fast, eltayeb2018compressive}.
In this case, the wireless channel can be simplified as the free-space transmission model, where $L_\mathrm{TX} = L_\mathrm{RX} = 1$ and all channel parameters in \eqref{chanModel} are available.
In this way, the channel can be obtained without channel training.

We simply assume that $\mathbf{h}_\mathrm{TX}$ and $\mathbf{h}_\mathrm{RX}$ are perfectly known for this case and develop the diagnostic technique in Sec. \ref{sec:caseIfCSI}.
The transmission model for this case is
\begin{equation}
\boxed{
\begin{aligned}
&\text{Case I (full CSI)}: \\
&{y = \sum_{n=1}^N \bigg[\underbrace{\mathbf{f} \circ \mathbf{h}_\mathrm{RX} \circ \mathbf{h}_\mathrm{TX}}_\text{known} \circ \underbrace{\mathbf{m}}_\text{unknown} \bigg]_n + w.}
\end{aligned}
}
\end{equation}

\subsubsection{Case II (Partial CSI)}
\label{sec:chanpCSI}
For outdoor online diagnosis, the channels are subject to inevitable multipath scattering, which makes the above channel calculation unreliable.
Fortunately, for some applications, the IRS is installed in a fixed location (e.g. walls, ceilings, etc.) and one can adopt a base station as the RX which is also fixed.
Hence, both the IRS and the RX are located in an open area and the channel $\mathbf{h}_\mathrm{RX}$ between them can be assumed as a long-term stationary multipath-free channel \cite{zhou2020robust}, which can then be calculated using \eqref{chanModel}.

In this case, we simply assume that $\mathbf{h}_\mathrm{RX}$ is perfectly known, while $L_\mathrm{TX} \ge 1$ due to the multipath scattering environment around the TX and the channel parameters associated with the TX are unknown.
Note that since the mmWave channel is sparse in the angular domain \cite{rappaport2013millimeter}, there are only a few dominant sub-paths in the channel and we have $L_\mathrm{TX} \ll \min(H,W)$.
We will develop the diagnostic technique for this case in Sec. \ref{sec:caseIIpCSI}.
The transmission model for this case is
\begin{equation}
\boxed{
\begin{aligned}
&\text{Case II (partial CSI)}: \\
&{y = \sum_{n=1}^N \bigg[ \underbrace{\mathbf{f} \circ \mathbf{h}_\mathrm{RX}}_\text{known} \circ \underbrace{\mathbf{h}_\mathrm{TX} \circ \mathbf{m}}_\text{unknown}\bigg]_n + w.}
\end{aligned}
}
\end{equation}

\subsubsection{Case III (No CSI)}
Finally, we consider the most general case where both channels are subject to multipath scattering and thus all channel parameters are assumed unknown, except for the assumption of sparse mmWave channel.
We will develop the diagnostic technique for this case in Sec. \ref{sec:caseIIInCSI}.
The transmission model for this case is
\begin{equation}
\boxed{
\begin{aligned}
&\text{Case III (no CSI)}: \\
&{y = \sum_{n=1}^N \bigg[ \underbrace{\mathbf{f}}_\text{known} \circ \underbrace{\mathbf{h}_\mathrm{RX} \circ \mathbf{h}_\mathrm{TX} \circ \mathbf{m}}_\text{unknown}\bigg]_n + w.}
\end{aligned}
}
\end{equation}

\section{Case I: Diagnosis with Full CSI}
\label{sec:caseIfCSI}
In the case of full CSI, we assume both channels are known, i.e.,
\begin{equation}
\label{sysModelAll_case1}
y = \sum_{n=1}^N \bigg[ \underbrace{\mathbf{f} \circ \mathbf{h}_\mathrm{RX} \circ \mathbf{h}_\mathrm{TX}}_\text{known} \circ \underbrace{\mathbf{m}}_\text{unknown}\bigg]_n + w.
\end{equation}
Therefore, we can locally generate fault-free reference signal using known channel vectors.
By subtracting the reference signal from the received signal, the impact of channel can be eliminated.
The diagnosis problem is then formulated as a compressed sensing problem by exploiting the failure sparsity, which requires far fewer measurements than the total number of reflecting elements.

\subsection{Problem Formulation}
Absorbing known channel vectors into the weighing vector $\mathbf{f}$, the transmission model \eqref{sysModelAll_case1} can be rewritten as
\begin{equation}
\begin{aligned}
y 
&= \sum_{n=1}^N \big[ \mathbf{f} \circ \mathbf{h}_\mathrm{RX} \circ \mathbf{h}_\mathrm{TX} \circ \mathbf{m}\big]_n + w \\
&= \big( \underbrace{\mathbf{f} \circ \mathbf{h}_\mathrm{RX} \circ \mathbf{h}_\mathrm{TX}}_{=\mathbf{f}_\mathrm{h}} \big)^\mathrm{T} \mathbf{m} + w,
\end{aligned}
\end{equation}
where $\mathbf{f}_\mathrm{h} = \mathbf{f} \circ \mathbf{h}_\mathrm{RX} \circ \mathbf{h}_\mathrm{TX}$ is the equivalent sensing vector.

During $K$ time slots within the channel coherence time, the IRS uses different weighting vectors to reflect the signal, yielding the following observation equation:
\begin{equation}
\label{fCSImodel_ls}
\mathbf{y} = \mathbf{F}_\mathrm{h} \mathbf{m} + \mathbf{w},
\end{equation}
where $\mathbf{y} = [y_1, \cdots, y_K]^\mathrm{T}$, $[\mathbf{F}_\mathrm{h}]_{k,:} = \mathbf{f}_{\mathrm{h},k}^\mathrm{T}$, and $\mathbf{w} = [w_1, \cdots, w_K]^\mathrm{T}$, in which $y_k = \mathbf{f}_{\mathrm{h},k}^\mathrm{T} \mathbf{m} + w_k$, $\mathbf{f}_{\mathrm{h},k}$, and $w_k$ are the measured symbol, the sensing vector, and the noise associated with the $k$-th measurement, respectively.

Recall that the goal of diagnosis is to recover the failure mask vector $\mathbf{m}$ from the received signal $\mathbf{y}$.
To recover $\mathbf{m}$ under the linear observation model \eqref{fCSImodel_ls}, we can exploit the failure sparsity and turn \eqref{fCSImodel_ls} into a compressed sensing model.

Since $\mathbf{h}_\mathrm{RX}$ and $\mathbf{h}_\mathrm{TX}$ are known, the RX can locally generate the fault-free reference signal as
\begin{equation}
\mathbf{y}^\mathrm{r} = \mathbf{F}_\mathrm{h} \mathbf{1}_{N \times 1}.
\end{equation}
Subtracting the fault-free reference signal $\mathbf{y}^\mathrm{r}$ from the received signal $\mathbf{y}$, we can obtain the differential signal
\begin{equation}
\begin{aligned}
\mathbf{y}^\mathrm{d} 
&= \mathbf{y} - \mathbf{y}^\mathrm{r} \\
&= \mathbf{F}_\mathrm{h} \mathbf{m} - \mathbf{F}_\mathrm{h} \mathbf{1}_{N \times 1} + \mathbf{w} \\
&= \mathbf{F}_\mathrm{h} (\underbrace{\mathbf{m} - \mathbf{1}_{N \times 1}}_{=\mathbf{x}}) + \mathbf{w}.
\end{aligned}
\end{equation}
Due to the failure sparsity that most entries in $\mathbf{m}$ are unity, $\mathbf{x} = \mathbf{m} - \mathbf{1}_{N \times 1}$ is a sparse vector.
Therefore, we can recover $\mathbf{x}$ from compressed measurements (i.e., $K \ll N$) and then retrieve $\mathbf{m}$, in which non-unity entries indicate faulty reflecting elements.\footnote{
Due to noise, hardware imperfections, and limited measurements, the recovered $\mathbf{x} = \mathbf{m} - \mathbf{1}_{N \times 1}$ may not be strictly sparse. One can use thresholding to identify faulty reflecting elements. For example, elements satisfying $|\hat{m}_n - 1|\ge \mathrm{th}$ can be identified as faulty, where $\mathrm{th}$ is a threshold and can be set to the maximum tolerable failure intensity. The diagnostic techniques proposed in Sec. \ref{sec:caseIIpCSI} and \ref{sec:caseIIInCSI} can follow the same procedure to identify faulty reflecting elements.
}

\subsection{Failure Mask Recovery}
We propose to solve the following LASSO problem to recover $\mathbf{x}$ \cite{boyd2011distributed}:
\begin{equation}
\label{LASSOpro}
\hat{\mathbf{x}} = \arg \min_{\mathbf{x}} \Vert \mathbf{y}^\mathrm{d} - \mathbf{F}_\mathrm{h} \mathbf{x} \Vert_2^2 + \lambda_1 \Vert \mathbf{x} \Vert_1,
\end{equation}
where $l_2$ and $l_1$ norms represent the approximation error and the penalty on sparsity, respectively, and $\lambda_1$ is a regularization parameter controlling the balance between them.
Once $\mathbf{x}$ is recovered, we can retrieve $\mathbf{m}$ by
\begin{equation}
\hat{\mathbf{m}} = \hat{\mathbf{x}} + \mathbf{1}_{N \times 1}.
\end{equation}

If each element of the weighting vector $\mathbf{f}$ is drawn uniformly randomly from a set containing all possible weighing coefficients, low mutual correlation of $\mathbf{F}_\mathrm{h}$ can be achieved and thus $\mathbf{m}$ can be accurately recovered from about $s \log N$ random incoherent measurements with high probability \cite{candes2011probabilistic}, where $s$ is the number of faulty reflecting elements (i.e., the sparsity of $\mathbf{x}$).
Therefore, the diagnostic time can be significantly shortened by exploiting the failure sparsity.

The computational complexity of the diagnosis with full CSI is dominated by the $N$-dimensional matrix inversion in the algorithm for solving \eqref{LASSOpro} \cite{boyd2011distributed}, and thus the overall computational complexity is $O(N^3)$.

\section{Case II: Diagnosis with Partial CSI}
\label{sec:caseIIpCSI}
In the case of partial CSI, we assume that only $\mathbf{h}_\mathrm{RX}$ is known and $\mathbf{h}_\mathrm{TX}$ is unavailable, i.e.,
\begin{equation}
\label{sysModelAll_case2}
y = \sum_{n=1}^N \bigg[ \underbrace{\mathbf{f} \circ \mathbf{h}_\mathrm{RX}}_\text{known} \circ \underbrace{\mathbf{h}_\mathrm{TX} \circ \mathbf{m}}_\text{unknown}\bigg]_n + w.
\end{equation}
Therefore, both channel vector $\mathbf{h}_\mathrm{TX}$ and failure mask $\mathbf{m}$ need to be estimated, which can be challenging since they are coupled.
In this section, we jointly exploit the sparsities of mmWave channel and failure, and formulate the diagnosis problem as a cSLRMR problem.

\subsection{Problem Formulation}
The transmission model \eqref{sysModelAll_case2} can be reformulated as
\begin{equation}
\label{matTransModel}
\begin{aligned}
y 
&= \sum_{n=1}^N \big[ \mathbf{f} \circ \mathbf{h}_\mathrm{RX} \circ \mathbf{h}_\mathrm{TX} \circ \mathbf{m}\big]_n + w \\
&= \sum_{i,j} \left[\mathbf{H}_\mathrm{RX} \circ \mathbf{F}_\mathrm{mat} \circ \mathbf{M}_\mathrm{mat} \circ \mathbf{H}_\mathrm{TX}\right]_{i,j}+ w,
\end{aligned}
\end{equation}
where $\mathbf{F}_\mathrm{mat} \in \mathbb{C}^{H \times W}$ and $\mathbf{M}_\mathrm{mat} \in \mathbb{C}^{H \times W}$ are the matrices reshaped from $\mathbf{f}$ and $\mathbf{m}$, respectively, i.e., $\mathrm{vec}(\mathbf{F}_\mathrm{mat}) = \mathbf{f}$ and $\mathrm{vec}(\mathbf{M}_\mathrm{mat}) = \mathbf{m}$;
$\sum_{i,j}$ denotes the summation over all entries in the matrix.

The known channel matrix $\mathbf{H}_\mathrm{RX}$ can be absorbed into the weighting matrix $\mathbf{F}_\mathrm{mat}$ as
\begin{equation}
y = \sum_{i,j} \left[\mathbf{F}_\mathrm{RX} \circ \mathbf{M}_\mathrm{mat} \circ \mathbf{H}_\mathrm{TX}\right]_{i,j} + w,
\end{equation}
where $\mathbf{F}_\mathrm{RX}=\mathbf{H}_\mathrm{RX} \circ \mathbf{F}_\mathrm{mat}$.

In addition, the multiplicative failure mask $\mathbf{M}_\mathrm{mat}$ can be equivalently replaced by an additive failure deviation $\mathbf{D}_\mathrm{mat}$ as
\begin{equation}
\label{SLRMDmeaModel}
y = \sum_{i,j} \bigg[\mathbf{F}_\mathrm{RX} \circ (\underbrace{\mathbf{D}_\mathrm{mat} + \mathbf{H}_\mathrm{TX}}_{=\mathbf{X}})\bigg]_{i,j} + w,
\end{equation}
where $\mathbf{D}_\mathrm{mat} = \mathbf{H}_\mathrm{TX} \circ (\mathbf{M}_\mathrm{mat} - \mathbf{1}_{H \times W})$.

Equation \eqref{SLRMDmeaModel} is a linear observation model where $\mathbf{F}_\mathrm{RX}$ and $\mathbf{X}$ are the configurable weighting matrix and the signal to be estimated, respectively, and we denote it as $y = \mathcal{F}_{\mathbf{F}_\mathrm{RX}}(\mathbf{X})$.
We use $K$ different weighting matrices to measure the signal $\mathbf{X}$ and obtain the observation equation
\begin{equation}
	\label{SLmat}
	\mathbf{y} = \mathcal{F}(\mathbf{X}) + \mathbf{w},
\end{equation}
where $\mathbf{y} = [y_1, \cdots, y_K]^\mathrm{T}$ and $\mathbf{w} = [w_1, \cdots, w_K]^\mathrm{T}$;
$y_k = \mathcal{F}_{\mathbf{F}_{\mathrm{RX},k}}(\mathbf{X}) + w_k$, $\mathbf{F}_{\mathrm{RX},k}$, and $w_k$ are the measured symbol, the weighting matrix, and the noise associated with the $k$-th measurement, respectively.

The goal of diagnosis is to recover $\mathbf{D}_\mathrm{mat}$ and $\mathbf{H}_\mathrm{TX}$ simultaneously, and thus the failure mask $\mathbf{M}_\mathrm{mat}$ can be retrieved from $\mathbf{M}_\mathrm{mat} = \mathbf{D}_\mathrm{mat} \oslash \mathbf{H}_\mathrm{TX} + \mathbf{1}_{H \times W}$, in which non-unity entries indicate faulty reflecting elements.
However, the channel and the failure deviation are coupled since both of them are unknown.
To recover $\mathbf{H}_\mathrm{TX}$ and $\mathbf{D}_\mathrm{mat}$ simultaneously, we need to introduce \textit{a-prior} information on them.
Note that $\mathbf{D}_\mathrm{mat} = \mathbf{H}_\mathrm{TX} \circ (\mathbf{M}_\mathrm{mat} - \mathbf{1}_{H \times W})$ is a sparse matrix due to the failure sparsity that most entries in $\mathbf{M}_\mathrm{mat}$ are unity.
According to the channel assumption in Sec. \ref{sec:chanpCSI} (sparse mmWave channel), $\mathbf{H}_\mathrm{TX}$ is a low-rank matrix since $\mathrm{rank}(\mathbf{H}_\mathrm{TX}) \le L_\mathrm{TX} \ll \min(H,W)$.
Therefore, $\mathbf{X}=\mathbf{D}_\mathrm{mat} + \mathbf{H}_\mathrm{TX}$ is the sum of a sparse and a low-rank matrix, which allows $\mathbf{H}_\mathrm{TX}$ and $\mathbf{D}_\mathrm{mat}$ to be recovered simultaneously from compressed measurements thanks to the rank-sparsity incoherence \cite{candes2011robust, wright2013compressive}.

\subsection{Failure Mask Recovery}
One can solve the following optimization problem to recover $\mathbf{H}_\mathrm{TX}$ and $\mathbf{D}_\mathrm{mat}$ \cite{tanner2020compressed}:
\begin{equation}
\label{P2}
\begin{aligned}
\{ \hat{\mathbf{H}}_\mathrm{TX}, \hat{\mathbf{D}}_\mathrm{mat} \} 
=& \arg\min_{\mathbf{H}_\mathrm{TX}, \mathbf{D}_\mathrm{mat}} \Vert \mathbf{H}_\mathrm{TX} \Vert_* + \lambda_2 \Vert \mathbf{D}_\mathrm{mat} \Vert_1 \\
&\mathrm{s.t.} \hspace{0.5em} \Vert \mathbf{y}  - \mathcal{F} (\mathbf{D}_\mathrm{mat} + \mathbf{H}_\mathrm{TX})\Vert_2 \le \delta,
\end{aligned}
\end{equation}
where $\Vert \cdot \Vert_*$ denotes the nuclear norm (sum of singular values) and represents the penalty on the rank;
$\lambda_2$ is a regularization parameter controlling the balance between the low-rank component and the sparse component;
$\delta$ is the noise level and is related to the SNR.
The optimization problem \eqref{P2} is convex and can be solved efficiently using the algorithm proposed in \cite{tanner2020compressed}.
Once $\mathbf{H}_\mathrm{TX}$ and $\mathbf{D}_\mathrm{mat}$ are recovered, the failure mask $\mathbf{m}$ can be retrieved by
\begin{equation}
\hat{\mathbf{m}} = \mathrm{vec}\left(\hat{\mathbf{M}}_\mathrm{mat}\right) = \mathrm{vec} \left( \hat{\mathbf{D}}_\mathrm{mat} \oslash \hat{\mathbf{H}}_\mathrm{TX} + \mathbf{1}_{H \times W} \right),
\end{equation}
where $\oslash$ denotes Hadamard division (element-wise division).

The computational complexity of the diagnosis with partial CSI is difficult to analyze since \eqref{P2} is eventually solved by the internal algorithms in the CVX toolbox.
To evaluate the computational efficiency in an alternative way, we provide the running time performance of the diagnosis with partial CSI in Sec. \ref{sec:simu}.

\section{Case III: Diagnosis without CSI}
\label{sec:caseIIInCSI}
In the case of no CSI, we assume both $\mathbf{h}_\mathrm{TX}$ and $\mathbf{h}_\mathrm{RX}$ are unavailable, i.e.,
\begin{equation}
\label{sysModelAll_case3}
y = \sum_{n=1}^N \bigg[ \underbrace{\mathbf{f}}_\text{known} \circ \underbrace{\mathbf{h}_\mathrm{RX} \circ \mathbf{h}_\mathrm{TX} \circ \mathbf{m}}_\text{unknown}\bigg]_n + w.
\end{equation}
This is the most challenging case since both channels are coupled with the failure mask.
In this section, we introduce a novel atomic norm to induce the sparsity of the cascaded channel $\mathbf{h}_\mathrm{RX} \circ \mathbf{h}_\mathrm{TX}$ and propose an ANM-based diagnostic technique to jointly recover the cascaded channel and the failure mask.
In addition, we extend the proposed diagnostic technique to the multi-antenna setting.
Two efficient algorithms are also developed to accelerate the diagnosis.

\subsection{Problem Formulation}
The transmission model \eqref{sysModelAll_case3} can be rewritten as
\begin{equation}
\label{ANMfirst}
\begin{aligned}
y 
&= \sum_{n=1}^N \bigg[ \mathbf{f} \circ \underbrace{\mathbf{h}_\mathrm{RX} \circ \mathbf{h}_\mathrm{TX}}_\text{cascaded channel $\mathbf{h}$} \circ \mathbf{m}\bigg]_n + w \\
&= \mathbf{f}^\mathrm{T}(\mathbf{h} \circ \mathbf{m}) + w \\
&= \mathbf{f}^\mathrm{T}(\mathbf{h} + \mathbf{d}) + w,
\end{aligned}
\end{equation}
where $\mathbf{h} = \mathbf{h}_\mathrm{TX} \circ \mathbf{h}_\mathrm{RX}$ is the cascaded channel and $\mathbf{d} = \mathbf{h} \circ (\mathbf{m} - \mathbf{1}_{N \times 1})$ is the additive failure deviation.

Similarly, after $K$ measurements, we have the observation equation
\begin{equation}
\label{SMV}
\mathbf{y} = \mathbf{F} (\mathbf{h+d}) + \mathbf{w},
\end{equation}
where $\mathbf{y} = [y_1, \cdots, y_K]^\mathrm{T}$, $[\mathbf{F}]_{k,:} = \mathbf{f}_k^\mathrm{T}$, and $\mathbf{w} = [w_1, \cdots, w_K]^\mathrm{T}$, in which $y_k = \mathbf{f}_k^\mathrm{T} (\mathbf{h+d}) + w_k$, $\mathbf{f}_k$, and $w_k$ are the measured symbol, the sensing vector, and the noise associated with the $k$-th measurement, respectively.

Following a similar idea to that in Sec. \ref{sec:caseIIpCSI}, we need to introduce \textit{a-prior} information on the unknown cascaded channel $\mathbf{h}$ to decouple it from the failure deviation $\mathbf{d}$.
One possible approach is to use the nuclear norm as adopted in the previously proposed diagnostic technique for Case II.
However, the cascaded channel may no longer be low-rank since $\mathrm{rank}(\mathbf{H}_\mathrm{RX} \circ \mathbf{H}_\mathrm{TX}) \le \mathrm{rank}(\mathbf{H}_\mathrm{RX}) \mathrm{rank}(\mathbf{H}_\mathrm{TX}) \le L_\mathrm{RX} L_\mathrm{TX}$, which means that even moderate multipath scattering may lead to a full-rank cascaded channel.
To cope with this issue, in the following, we introduce a novel atomic norm as the sparsity-inducing norm of the cascaded channel.

\subsection{The Atomic Norm of Cascaded Channel}
Using the Kronecker product form of channel model \eqref{chanModel}, the cascaded channel $\mathbf{h}$ can be expressed as
\begin{equation}
\begin{aligned}
\mathbf{h}
=& \mathbf{h}_\mathrm{TX} \circ \mathbf{h}_\mathrm{RX} \\
=& \left( \sum_{l=1}^{L_\mathrm{TX}} \alpha_{\mathrm{TX},l} \bigg( \mathbf{a}_\mathrm{W}(\theta_{\mathrm{TX},l},\phi_{\mathrm{TX},l}) \otimes \mathbf{a}_\mathrm{H}(\theta_{\mathrm{TX},l},\phi_{\mathrm{TX},l}) \bigg) \right) \\
& \circ
\left( \sum_{l=1}^{L_\mathrm{RX}} \alpha_{\mathrm{RX},l} \bigg( \mathbf{a}_\mathrm{W}(\theta_{\mathrm{RX},l},\phi_{\mathrm{RX},l}) \otimes \mathbf{a}_\mathrm{H}(\theta_{\mathrm{RX},l},\phi_{\mathrm{RX},l}) \bigg) \right) \\
=& \sum_{i=1}^{L_\mathrm{TX}} \sum_{j=1}^{L_\mathrm{RX}} \alpha_{\mathrm{TX},i} \alpha_{\mathrm{RX},j} \bigg( \mathbf{a}_\mathrm{W}(\theta_{\mathrm{TX},i},\phi_{\mathrm{TX},i}) \otimes \mathbf{a}_\mathrm{H}(\theta_{\mathrm{TX},i},\phi_{\mathrm{TX},i}) \bigg) \\
& \circ \bigg( \mathbf{a}_\mathrm{W}(\theta_{\mathrm{RX},j},\phi_{\mathrm{RX},j}) \otimes \mathbf{a}_\mathrm{H}(\theta_{\mathrm{RX},j},\phi_{\mathrm{RX},j}) \bigg) \\
=& \sum_{i=1}^{L_\mathrm{TX}} \sum_{j=1}^{L_\mathrm{RX}} \alpha_{\mathrm{TX},i} \alpha_{\mathrm{RX},j} \bigg( \mathbf{a}_\mathrm{W}(\theta_{\mathrm{TX},i},\phi_{\mathrm{TX},i}) \circ \mathbf{a}_\mathrm{W}(\theta_{\mathrm{RX},j},\phi_{\mathrm{RX},j})  \bigg) \\
& \otimes \bigg( \mathbf{a}_\mathrm{H}(\theta_{\mathrm{TX},i},\phi_{\mathrm{TX},i}) \circ \mathbf{a}_\mathrm{H}(\theta_{\mathrm{RX},j},\phi_{\mathrm{RX},j}) \bigg) \\
=& \sum_{i=1}^{L_\mathrm{TX}} \sum_{j=1}^{L_\mathrm{RX}} \alpha_{\mathrm{TX},i} \alpha_{\mathrm{RX},j} \bigg( \mathbf{a}_\mathrm{W}'(\theta_{\mathrm{TX},i}, \phi_{\mathrm{TX},i}, \theta_{\mathrm{RX},j}, \phi_{\mathrm{RX},j}) \\
& \otimes \mathbf{a}_\mathrm{H}'(\theta_{\mathrm{TX},i}, \phi_{\mathrm{TX},i}, \theta_{\mathrm{RX},j}, \phi_{\mathrm{RX},j}) \bigg),
\end{aligned}
\end{equation}
where
\begin{equation}
\begin{aligned}
\mathbf{a}_\mathrm{W}'(\theta_i, \phi_i, \theta_j, \phi_j)
=& \mathbf{a}_\mathrm{W}(\theta_i,\phi_i) \circ \mathbf{a}_\mathrm{W}(\theta_j,\phi_j) \\
=& [1, e^{j2\pi d (\sin \theta_i \cos \phi_i + \sin \theta_j \cos \phi_j)}, \cdots, \\
& e^{j2\pi d (W-1) (\sin \theta_i \cos \phi_i + \sin \theta_j \cos \phi_j)}]^\mathrm{T}, \\
\mathbf{a}_\mathrm{H}'(\theta_i, \phi_i, \theta_j, \phi_j)
=& \mathbf{a}_\mathrm{H}(\theta_i,\phi_i) \circ \mathbf{a}_\mathrm{H}(\theta_j,\phi_j) \\
=& [1, e^{j2\pi d (\sin \theta_i \sin \phi_i + \sin \theta_j \sin \phi_j)}, \cdots, \\
& e^{j2\pi d (H-1) (\sin \theta_i \sin \phi_i + \sin \theta_j \sin \phi_j)}]^\mathrm{T}.
\end{aligned}
\end{equation}
Therefore, the cascaded channel can be treated as a channel that has $L = L_\mathrm{TX} L_\mathrm{RX}$ equivalent sub-paths.
To lighten the notation, we rewrite the cascaded channel $\mathbf{h}$ as
\begin{equation}
\mathbf{h} = \sum_{l=1}^{L} \alpha_l \mathbf{a}_\mathrm{W}(g_{\mathrm{W},l}) \otimes \mathbf{a}_\mathrm{H}(g_{\mathrm{H},l}) = \sum_{l=1}^{L} \alpha_l \mathbf{a}(\mathbf{g}_l),
\end{equation}
where 
\begin{equation}
\begin{aligned}
\mathbf{a}(\mathbf{g}_l) &= \mathbf{a}_\mathrm{W}(g_{\mathrm{W},l}) \otimes \mathbf{a}_\mathrm{H}(g_{\mathrm{H},l}), \\
\mathbf{a}_\mathrm{W}(g_{\mathrm{W},l}) &= [1, e^{j2\pi d g_{\mathrm{W},l}}, \cdots, e^{j2\pi d (W-1) g_{\mathrm{W},l}}]^\mathrm{T}, \\
\mathbf{a}_\mathrm{H}(g_{\mathrm{H},l}) &= [1, e^{j2\pi d g_{\mathrm{H},l}}, \cdots, e^{j2\pi d (H-1) g_{\mathrm{H},l}}]^\mathrm{T},
\end{aligned}
\end{equation}
and
$\alpha_l \in \{ \alpha_{\mathrm{TX},i} \alpha_{\mathrm{RX},j}, (i=1,\cdots,L_\mathrm{TX},
j=1\cdots,L_\mathrm{RX}) \}$,
$\mathbf{g}_l = [g_{\mathrm{W},l}, g_{\mathrm{H},l}]^\mathrm{T}$,
$g_{\mathrm{W},l} \in \{ \sin \theta_i \cos \phi_i + \sin \theta_j \cos \phi_j, (i=1,\cdots,L_\mathrm{TX}, j=1\cdots,L_\mathrm{RX}) \}$,
$g_{\mathrm{H},l} \in \{ \sin \theta_i \sin \phi_i + \sin \theta_j \sin \phi_j, (i=1,\cdots,L_\mathrm{TX}, j=1\cdots,L_\mathrm{RX}) \}$.

From the above derivation, we can observe that the cascaded channel $\mathbf{h}$ is the sum of $L$ 2D frequency components, which has a sparse representation in the 2D Fourier dictionary since $L\ll N$.
To represent $\mathbf{h}$ in a sparse form, a conventional approach is to discretize the angular domain into a grid and form a 2D discrete Fourier transform (DFT) dictionary \cite{yang2012off}.
However, this approach incurs off-grid error since frequency components (AOAs/AoDs) do not lie exactly on the grid, and hence the performance of sparse recovery algorithms may degrade considerably \cite{chi2011sensitivity}.
Although increasing the grid density can alleviate the off-grid problem to a certain extent, the computational complexity also increases dramatically, especially under the 2D setting.
To bypass this issue, we adopt the atomic norm as the sparsity promoter of the cascaded channel, which does not discretize the angular domain and thus no off-grid error will be induced.
A comprehensive survey on the atomic norm can be found in \cite{chi2020harnessing}.

Recall that the cascaded channel $\mathbf{h}$ has a sparse representation in the 2D Fourier dictionary.
To be more specific, $\mathbf{h}$ is sparse in the continuous 2D Fourier dictionary\footnote{
Strictly speaking, $\mathbf{g}$ lies in the range $[-2,2) \times [-2,2)$. However, due to the periodic property of the exponent, only one period (i.e., $[-1,1) \times [-1,1)$) is actually required.}
\begin{equation}
\mathcal{A} = \{\mathbf{a}(\mathbf{g}) | \mathbf{g} \in [-1,1) \times [-1,1) \},
\end{equation}
which contains the building blocks (atoms) of the signal $\mathbf{h}$.
Analogous to the definition of $l_1$ norm, the atomic norm of $\mathbf{h}$ over the atom set $\mathcal{A}$ is defined as
\begin{equation}
\label{aNorm}
\Vert \mathbf{h} \Vert_\mathcal{A} = \inf \left\{ \sum_{l} |\alpha_l| \bigg| \mathbf{h} = \sum_l \alpha_l \mathbf{a}_l, \hspace{0.25em} \mathbf{a}_l \in \mathcal{A} \right \},
\end{equation}
which is the minimum cost (in terms of $l_1$ norm measure) of representing a sparse signal.

Although the definition of atomic norm is simple and clear, it can not be directly minimized since it involves the optimization of infinite-dimensional variables.
To minimize the atomic norm \eqref{aNorm} in a tractable manner, one can exploit the Vandermonde structure of signal $\mathbf{h}$ and solve the following positive semidefinite program \cite{chi2014compressive}:
\begin{equation}
\label{anappro}
\arg \min_{\substack{\mathbf{u} \in \mathbb{C}^{N_u} \\ v \in \mathbb{R}}}
\frac{1}{2}\left(\frac{1}{N} \mathrm{tr}\big(T(\mathbf{u})\big) + v \right)
\hspace{0.5em} \mathrm{s.t.} \hspace{0.5em}
\begin{bmatrix}
T(\mathbf{u}) & \mathbf{h} \\
\mathbf{h}^\mathrm{H} & v
\end{bmatrix}
\succeq 0,
\end{equation}
where $T(\mathbf{u})$ is a twofold Toeplitz operator, which maps a given vector
\begin{equation}
\begin{aligned}
\mathbf{u} 
=& [\underbrace{u_{0,0}, u_{0,1}, \cdots, u_{0,W-1}}_\text{block 0}, \underbrace{u_{1,-(W-1)}, \cdots, u_{1,W-1}}_\text{block 1}, \\
& \cdots \cdots, \underbrace{u_{H-1, -(W-1)}, \cdots, u_{H-1, W-1}}_\text{block $H-1$}]
\end{aligned}
\end{equation}
of length $N_u = (H-1)(2W-1)+W$ into a $H \times H$ block Toeplitz matrix
\begin{equation}
T(\mathbf{u}) = 
\begin{bmatrix}
\mathbf{T}_0 & \mathbf{T}_1^\mathrm{H} & \cdots & \mathbf{T}_{H-1}^\mathrm{H} \\
\mathbf{T}_1 & \mathbf{T}_0 & \cdots & \mathbf{T}_{H-2}^\mathrm{H} \\
\vdots & \vdots & \ddots & \vdots \\
\mathbf{T}_{H-1} & \mathbf{T}_{H-2} & \cdots & \mathbf{T}_0
\end{bmatrix},
\end{equation}
with each block $\mathbf{T}_i$ ($0 \le i < H$) being a $W \times W$ Toeplitz matrix
\begin{equation}
\mathbf{T}_i =
\begin{bmatrix}
u_{i,0} & u_{i,-1} & \cdots & u_{i,-(W-1)} \\
u_{i,1} & u_{i,0} & \cdots & u_{i,-(W-2)} \\
\vdots & \vdots & \ddots & \vdots \\
u_{i,W-1} & u_{i,W-2} & \cdots & u_{i,0} \\
\end{bmatrix}.
\end{equation}
Note that \eqref{anappro} is just an approximation for minimizing the atomic norm since there is a fundamental difficulty in generalizing the classical Caratheodory’s theorem to higher dimensions \cite{chi2014compressive}.
Nonetheless, \eqref{anappro} exhibits excellent empirical performance as illustrated later in numerical simulations (Sec. \ref{sec:simu}).

\subsection{Failure Mask Recovery}
Using the atomic norm and the $l_1$ norm as the sparsity promoters for the cascaded channel $\mathbf{h}$ and the failure deviation $\mathbf{d}$, respectively, we can simultaneously recover $\mathbf{h}$ and $\mathbf{d}$ by solving the following joint sparse recovery problem:
\begin{equation}
\label{ANMdirect}
\{ \hat{\mathbf{h}}, \hat{\mathbf{d}}\} = \arg \min_{\mathbf{h}, \mathbf{d}} \frac{1}{2} \Vert \mathbf{y-F(h + d)}\Vert_2^2 + \tau_1 \Vert \mathbf{h} \Vert_\mathcal{A} + \lambda_3 \Vert \mathbf{d} \Vert_1,
\end{equation} 
where $\tau_1$ and $\lambda_3$ are regularization parameters controlling the penalty on the sparsity of $\mathbf{h}$ and $\mathbf{d}$, respectively.

According to \eqref{anappro}, the optimization problem \eqref{ANMdirect} can be rewritten as
\begin{equation}
\label{ANMfinal}
\begin{aligned}
\{ \hat{\mathbf{h}}, \hat{\mathbf{d}}\}
=& \arg \min_{\mathbf{h, d, u}, v} \frac{1}{2}\Vert \mathbf{y-F(h + d)}\Vert_2^2 \\
& + \frac{\tau_1}{2} \left( \frac{1}{N} \mathrm{tr}\big(T(\mathbf{u})\big) + v \right)
+ \lambda_3 \Vert \mathbf{d} \Vert_1, \\
& \mathrm{s.t.}
\begin{bmatrix}
	T(\mathbf{u}) & \mathbf{h} \\
	\mathbf{h}^\mathrm{H} & v
\end{bmatrix}
\succeq 0.
\end{aligned}
\end{equation}
Once $\mathbf{h}$ and $\mathbf{d}$ are recovered, the failure mask $\mathbf{m}$ can be retrieved by
\begin{equation}
\hat{\mathbf{m}} = \hat{\mathbf{d}} \oslash \hat{\mathbf{h}} + \mathbf{1}_{N \times 1}.
\end{equation}

The optimization problem \eqref{ANMfinal} is a positive semidefinite program and can be solved by off-the-shelf solvers like SDPT3 in the CVX toolbox, as suggested in numerous related works on atomic norm.
SDPT3 is based on the interior point method, which tends to be extremely time-consuming even for medium-scale systems.
In Appendix \ref{sec:ADMM}, we develop an efficient algorithm for solving \eqref{ANMfinal}, which is based on the ADMM algorithm.
One can refer to \cite{boyd2011distributed} for a thorough survey of the ADMM algorithm.

\subsection{Extension to Multiple Antennas at RX}
\label{sec:multiRX}
Recall that in this case we assume that there may be multipath scattering in the IRS-RX channel, which can be exploited by a multi-antenna RX to provide diversity gain.
We consider a single antenna at the TX and $N_\mathrm{RX}$ antennas at the RX.
For simplicity, we assume that antennas at the RX form a uniform linear array (ULA).
Note that the following proposed diagnostic technique can be easily extended to other array structures like UPA.

With multiple antennas at the RX, the transmission model \eqref{sysModelAll} can be extended to
\begin{equation}
\mathbf{y} = \mathbf{H}'_\mathrm{RX} (\mathrm{diag}(\mathbf{f}) \mathrm{diag}(\mathbf{m})) \mathbf{h}_\mathrm{TX} + \mathbf{w},
\end{equation}
where $\mathbf{y} \in \mathbb{C}^{N_\mathrm{RX}}$ is the received signal, $\mathbf{w} \in \mathbb{C}^{N_\mathrm{RX}}$ is the noise.
$\mathbf{H}'_\mathrm{RX} \in \mathbb{C}^{N_\mathrm{RX} \times N}$ is the channel matrix between the RX and the IRS, which can be expressed as
\begin{equation}
\begin{aligned}
\mathbf{H}'_\mathrm{RX}
=& \sum_{l=1}^{L_\mathrm{RX}} \alpha_{\mathrm{RX},l} \mathbf{a}_\mathrm{RX}(\theta'_{\mathrm{RX},l}) \\
& \times \bigg( \mathbf{a}_\mathrm{W}(\theta_{\mathrm{RX},l},\phi_{\mathrm{RX},l}) \otimes \mathbf{a}_\mathrm{H}(\theta_{\mathrm{RX},l},\phi_{\mathrm{RX},l}) \bigg)^\mathrm{T},
\end{aligned}
\end{equation}
where
$
\mathbf{a}_\mathrm{RX}(\theta) = [1, e^{j2\pi d \sin\theta}, \cdots, e^{j2\pi d (N_\mathrm{RX} -1) \sin\theta}]^\mathrm{T}
$
and $\theta'_{\mathrm{RX},l}$ is the AoA of the $l$-th sub-path in the IRS-RX channel.
Note that each entry in $\mathbf{y}$ can be expressed in the form of \eqref{ANMfirst} (transmission model with single-antenna RX) with different channel vectors $\mathbf{h}_\mathrm{RX}$.
Denoting the received symbol and the channel vector associated with the $i$-th antenna at the RX as $y_i$ and $\mathbf{h}_{\mathrm{RX},i}$, respectively, we can represent $\mathbf{y}$ and $\mathbf{H}'_\mathrm{RX}$ as
\begin{equation}
\begin{aligned}
\mathbf{y} &= [y_1, \cdots, y_{N_\mathrm{RX}}]^\mathrm{T}, \\
\mathbf{H}'_\mathrm{RX} &= [\mathbf{h}_{\mathrm{RX},1}, \cdots, \mathbf{h}_{\mathrm{RX},N_\mathrm{RX}}]^\mathrm{T}.
\end{aligned}
\end{equation}

According to the observation equation \eqref{SMV}, after $K$ measurements, the $i$-th antenna at the RX observes a vector of measurements
\begin{equation}
\mathbf{y}_i = \mathbf{F} (\mathbf{h}_i + \mathbf{d}_i) + \mathbf{w}_i,
\end{equation}
where $\mathbf{h}_i = \mathbf{h}_\mathrm{TX} \circ \mathbf{h}_{\mathrm{RX},i}$, $\mathbf{d}_i = \mathbf{h}_i \circ (\mathbf{m} - \mathbf{1}_{N \times 1})$, and $\mathbf{w}_i$ are the cascaded channel, the failure deviation, and the measurement noise associated with the $i$-th antenna at the RX, respectively.

Stacking the measurement vectors from all RX antennas into a matrix, we obtain the following MMV model:
\begin{equation}
\mathbf{Y} = \mathbf{F} (\mathbf{H+D}) + \mathbf{W},
\end{equation}
where $\mathbf{Y} = [\mathbf{y}_1, \cdots, \mathbf{y}_{N_\mathrm{RX}}]$, $\mathbf{H} = [\mathbf{h}_1, \cdots, \mathbf{h}_{N_\mathrm{RX}}]$, $\mathbf{D} = [\mathbf{d}_1, \cdots, \mathbf{d}_{N_\mathrm{RX}}]$, and $\mathbf{W} = [\mathbf{w}_1, \cdots, \mathbf{w}_{N_\mathrm{RX}}]$.

Note that all column vectors in $\mathbf{H}$ share the same sparse support (i.e., location of non-zero components) since AoAs/AoDs are the same with respect to each receiving channel vector $\mathbf{h}_{\mathrm{RX},i}$.
Besides, due to multipath scattering in the IRS-RX channel, $\mathbf{H}$ is of rank $L_\mathrm{RX}$, which provides a diversity gain of order $L_\mathrm{RX}$.
The failure deviation $\mathbf{D}$ also reserves such diversity gain since it is associated with $\mathbf{H}$ as $\mathbf{d}_i = \mathbf{h}_i \circ (\mathbf{m} - \mathbf{1}_{N \times 1})$.
Therefore, we can exploit the diversity gain provided by multipath scattering in the IRS-RX channel to improve the recovery performance.

The MMV-form sparsity-inducing norms for the channel $\mathbf{H}$ and the failure deviation $\mathbf{D}$ are introduced as follows.
The MMV-form atomic norm is defined as \cite{li2015off}
\begin{equation}
	\label{aNormMMV}
	\Vert \mathbf{H} \Vert_{\mathcal{A}_\mathrm{MMV}} = \inf \left\{ \sum_{l} |\alpha_l| \bigg| \mathbf{H} = \sum_l \alpha_l \mathbf{A}_l, \hspace{0.25em} \mathbf{A}_l \in \mathcal{A}_\mathrm{MMV} \right \},
\end{equation}
where
\begin{equation}
\mathcal{A}_\mathrm{MMV} = \{\mathbf{a}(\mathbf{g}) \mathbf{b}^\mathrm{T} | \mathbf{g} \in [-1,1) \times [-1,1), \Vert \mathbf{b} \Vert_2 = 1 \}
\end{equation}
and $\mathbf{b} \in \mathbb{C}^{N_\mathrm{RX}}$.
Similar to the single measurement vector (SMV) model \eqref{anappro}, the MMV-form atomic norm of $\mathbf{H}$ can be minimized by solving the positive semidefinite program \cite{li2015off}
\begin{equation}
\label{anapproMMV}
\begin{aligned}
\arg \min_{\substack{\mathbf{u} \in \mathbb{C}^{N_u} \\ \mathbf{V} \in \mathbb{C}^{N_\mathrm{RX} \times N_\mathrm{RX}}}}&
\frac{1}{2}\left(\frac{1}{N} \mathrm{tr}\big(T(\mathbf{u})\big) + \mathrm{tr}(\mathbf{V}) \right) \\
&\mathrm{s.t.}
\begin{bmatrix}
T(\mathbf{u}) & \mathbf{H} \\
\mathbf{H}^\mathrm{H} & \mathbf{V}
\end{bmatrix}
\succeq 0.
\end{aligned}
\end{equation}
For the MMV-form sparsity-inducing norm of $\mathbf{D}$, we adopt the notation \cite{cotter2005sparse}
\begin{equation}
\label{MMVl1norm}
\Vert \mathbf{D} \Vert_{2,1} = \sum_{n=1}^{N} \Vert [\mathbf{D}]_{n,:} \Vert_2,
\end{equation}
where $[\mathbf{D}]_{n,:}$ denotes the $n$-th row of $\mathbf{D}$.
It can be observed that the MMV-form atomic norm $\Vert \cdot \Vert_{\mathcal{A}_\mathrm{MMV}}$ and sparsity-inducing norm $\Vert \cdot \Vert_{2,1}$ reduce exactly to the SMV-form atomic norm $\Vert \cdot \Vert_\mathcal{A}$ and $l_1$ norm $\Vert \cdot \Vert_1$ when the number of RX antennas is $N_\mathrm{RX}=1$, respectively.

Using the above sparsity-inducing norms, the optimization problem for jointly recovering $\mathbf{H}$ and $\mathbf{D}$ can be expressed as
\begin{equation}
\label{ANMdirectMMV}
\begin{aligned}
\{ \hat{\mathbf{H}}, \hat{\mathbf{D}}\} = \arg \min_{\mathbf{H}, \mathbf{D}} \hspace{0.5em} & \frac{1}{2} \Vert \mathbf{Y-F(H + D)}\Vert_\mathrm{F}^2 \\
&+ \tau_2 \Vert \mathbf{H} \Vert_{\mathcal{A}_\mathrm{MMV}} + \lambda_4 \Vert \mathbf{D} \Vert_{2,1},
\end{aligned}
\end{equation} 
which admits positive semidefinite program
\begin{equation}
\label{ANMfinalMMV}
\begin{aligned}
\{ \hat{\mathbf{H}}, \hat{\mathbf{D}}\} 
=& \arg \min_{\mathbf{H, D, u}, \mathbf{V}} \hspace{0.5em}
\frac{1}{2}\Vert \mathbf{Y-F(H + D)}\Vert_\mathrm{F}^2 \\
&+ \frac{\tau_2}{2} \left( \frac{1}{N} \mathrm{tr}\big(T(\mathbf{u})\big) + \mathrm{tr}(\mathbf{V}) \right) + \lambda_4 \Vert \mathbf{D} \Vert_{2,1}, \\
& \mathrm{s.t.} \hspace{0.5em}
\begin{bmatrix}
	T(\mathbf{u}) & \mathbf{H} \\
	\mathbf{H}^\mathrm{H} & \mathbf{V}
\end{bmatrix}
\succeq 0.
\end{aligned}
\end{equation}
Once $\mathbf{H}$ and $\mathbf{D}$ are recovered, the failure mask $\mathbf{m}$ can be retrieved by
\begin{equation}
\hat{\mathbf{m}} = \mathrm{mean} \left( \hat{\mathbf{D}} \oslash \hat{\mathbf{H}} + \mathbf{1}_{N \times N_\mathrm{RX}} \right),
\end{equation}
where $\mathrm{mean}(\cdot)$ denotes the mean of each row.

The optimization problem \eqref{ANMfinalMMV} also enjoys an efficient ADMM-based algorithm, which is summarized in Appendix \ref{sec:ADMM2}.

It is worth noting that no diversity gain can be obtained if the TX equips multiple antennas, since the signal impinging on the IRS is fixed, no matter how many antennas are at the TX.
We can also consider using different precoding matrices to transmit the test symbol.
However, it will lead to a cascaded channel (signal to be estimated) that varies with each measurement, which even makes the diagnosis completely failed.

\section{Numerical Simulations}
\label{sec:simu}
This section presents several illustrative instances and numerical simulations to demonstrate the performance of proposed diagnostic techniques.
The MATLAB codes for implementing all the proposed algorithms are available online.\footnote{https://github.com/DestinationSR/IRSdiagnosis.}

\subsection{Parameter Setting and Performance Metric}
Default simulation parameters are listed in Table \ref{simuParaSet}.
We adopt the $\mathrm{NMSE}$ (normalized mean squared error) of recovered failure masks as the performance metric, which is defined as
\begin{equation}
\mathrm{NMSE} = \frac{\Vert \mathbf{m} - \hat{\mathbf{m}} \Vert_2^2}{\Vert \mathbf{m} \Vert_2^2}.
\end{equation}
Besides, the measurement number shown in the figures is normalized by the number of reflecting elements $N$ (nominal signal dimension).

\begin{table}
	\small
	\centering
	\caption{Default simulation parameters.}
	\label{simuParaSet}
	\renewcommand\arraystretch{1.15}
	\begin{tabular}{p{2.45cm}<{\raggedright} p{1.65cm}<{\raggedright} p{3.45cm}<{\raggedright}}
		\toprule
		\textbf{Parameter} & \textbf{Symbol} & \textbf{Setting} \\
		\midrule
		Number of reflecting elements & $N = H \times W$ & $16 \times 16$ \\
		\hline
		Resolution of phase weighing & & 2-bit, $[1+1j, 1-1j, -1+1j, -1-1j]/\sqrt{2}$ \\
		\hline
		Distribution of sub-path gain & $\alpha_\mathrm{TX}$, $\alpha_\mathrm{RX}$ & $\alpha_\mathrm{TX} \sim \mathcal{CN}(0,1/L_\mathrm{TX})$, $\alpha_\mathrm{RX} \sim \mathcal{CN}(0,1/L_\mathrm{RX})$ \\
		\hline
		Distribution of sub-path AoA/AoD  & $\theta$, $\phi$ & $\theta \sim U(0,2\pi)$, $\phi \sim U(0,2\pi)$ \\
		\hline
		Distribution of failure mask &$m = \eta e^{j \kappa}$ & $\eta \sim U(0,1)$, $\kappa \sim U(0,2\pi)$ \\
		\hline
		Regularization Parameters & $\lambda_1$, $\lambda_2$, $\lambda_3$, $\lambda_4$ & $\lambda_1=0.65K/\mathrm{SNR}$, $\lambda_2=0.35$, $\lambda_3=0.006K$, $\lambda_4=0.006K/N_\mathrm{RX}$ \\
		 & $\tau_1$, $\tau_2$ & $\tau_1=0.004K$, $\tau_2=0.004K$ \\
		\bottomrule
	\end{tabular}
\end{table}

\subsection{Illustrative Instances}
\begin{figure}
	\centering
	\subfigure[Diagnosis with full CSI, where red squares indicate the location of faulty reflecting elements. The impact of channel can be eliminated by comparing the received signal with the known reference signal, and thus the differential signal is a sparse signal that can be recovered from compressed measurements.]{
		\begin{minipage}[t]{0.95\linewidth}
			\centering
			\includegraphics[scale=0.6]{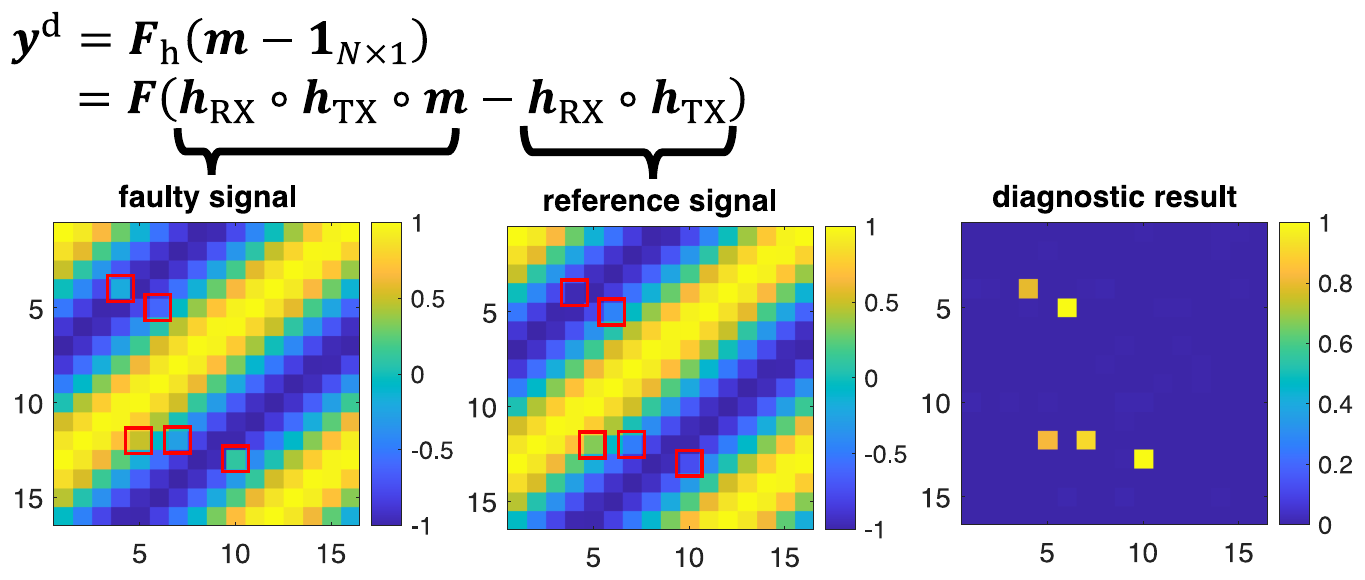}
			\label{fCSI_ins}
	\end{minipage}}
	
	\subfigure[Diagnosis with partial CSI. Although no reference signal is available, we can jointly exploit the low-rank property of the channel $\mathbf{H}_\mathrm{TX}$ and the sparsity of the failure deviation $\mathbf{D}$ to recover them simultaneously.]{
		\begin{minipage}[t]{0.95\linewidth}
			\centering
			\includegraphics[scale=0.6]{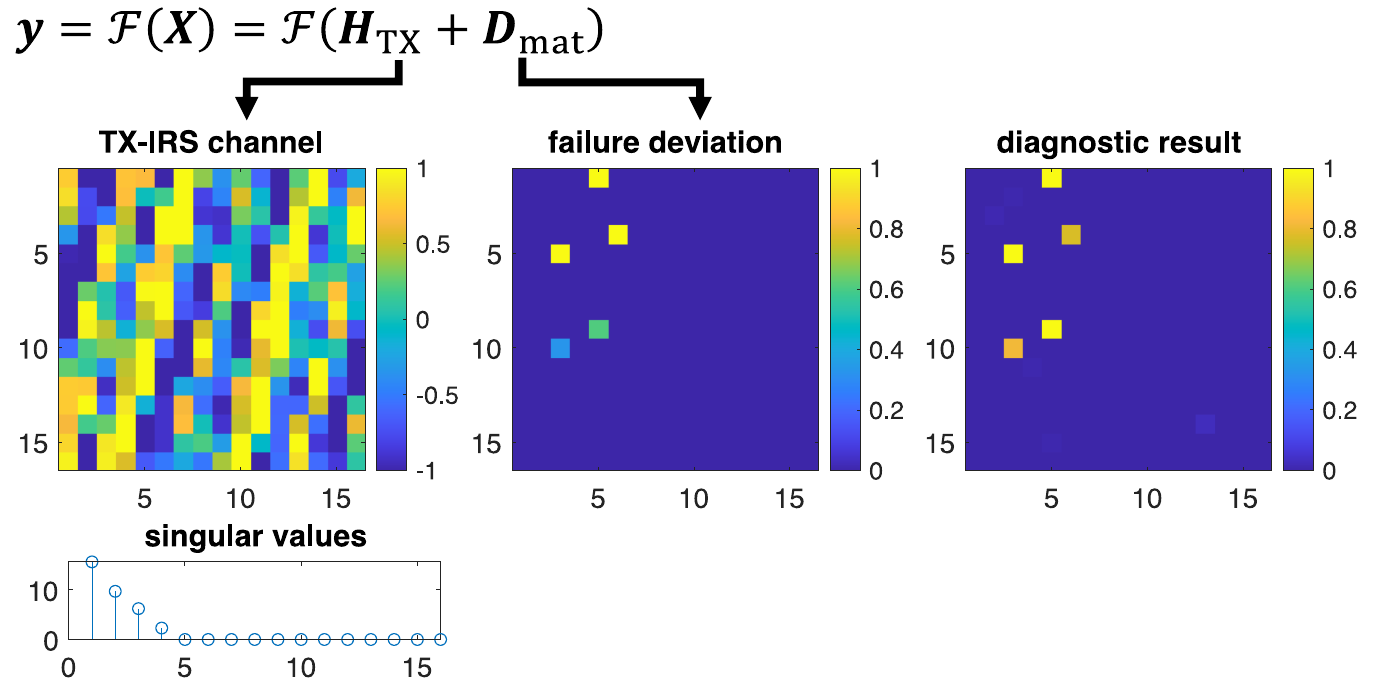}
			\label{pCSI_ins}
	\end{minipage}}
	
	\subfigure[Diagnosis without CSI. The cascaded channel $\mathbf{h}$ is not low-rank anymore, but it can be sparsely represented in the 2D Fourier dictionary. Using the atomic norm as the sparsity promoter, both cascaded channel and failure mask can be recovered.]{
		\begin{minipage}[t]{0.95\linewidth}
			\centering
			\includegraphics[scale=0.6]{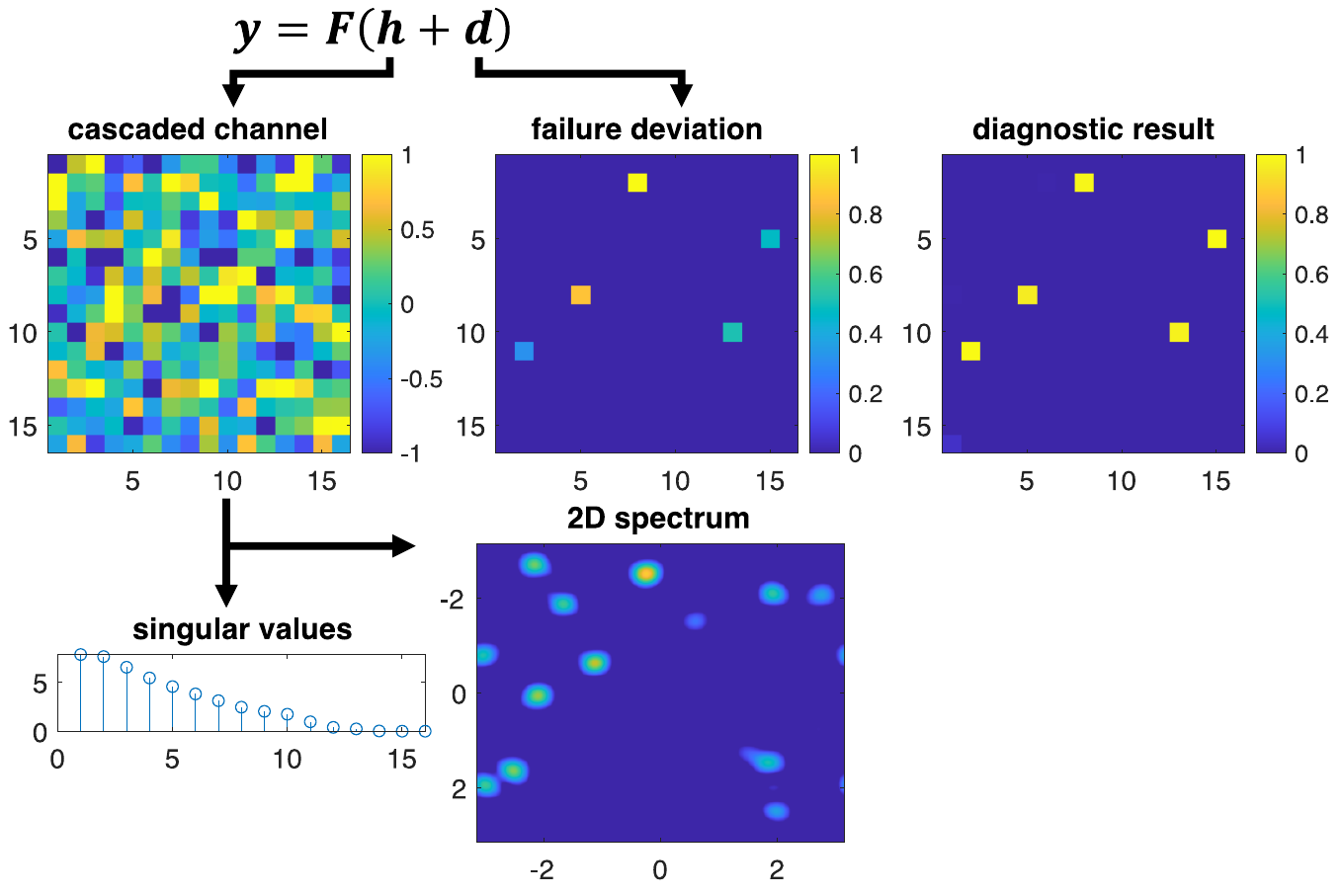}
			\label{nCSI_ins}
	\end{minipage}}
	
	\caption{Instances of proposed diagnostic techniques.}
	\label{instances}
\end{figure}

We first provide three instances in Fig. \ref{instances} to illustrate the main ideas of our proposed diagnostic techniques, where SNR$=20$ dB and $K=0.8N$.
For better visualization, all vectors are reshaped into matrices with size $H \times W$ in order to keep them consistent with the array structure.
The diagnostic results are presented in the form of failure intensity, which is defined as $\vert m_n - 1 \vert$.
Besides, only the real part of the channel is displayed.

The instance of the diagnosis with full CSI is shown in Fig. \ref{fCSI_ins}.
The main idea of the diagnosis with full CSI is to compare the received signal with the reference signal generated from known channel vectors.
Therefore, non-zero entries in the sparse differential signal indicate faulty reflecting elements, which can be recovered from compressed measurements.

The instance of the diagnosis with partial CSI is shown in Fig. \ref{pCSI_ins}, where the number of sub-paths in the TX-IRS channel is set to $L_\mathrm{TX} = 4$.
In the case of partial CSI, the channel fading and the failure are coupled and hence difficult to distinguish.
Note that the unknown channel matrix is low-rank since the number of sub-paths is limited.
Therefore, by jointly exploiting the channel low-rank property and the failure sparsity, $\mathbf{H}_\mathrm{TX}$ and $\mathbf{D}$ can be recovered simultaneously using the cSLRMR algorithm.

The instance of the diagnosis without CSI is shown in Fig. \ref{nCSI_ins}, where the number of sub-paths is $L_\mathrm{TX} = L_\mathrm{RX} = 4$.
We can see that the cascaded channel is not low-rank anymore since $\mathrm{rank}(\mathbf{H}_\mathrm{RX} \circ \mathbf{H}_\mathrm{TX}) = L_\mathrm{RX} L_\mathrm{TX} = 16$, while the 2D spectrum indicates that the cascaded channel can be sparsely represented in the 2D Fourier dictionary.
Therefore, using the atomic norm as the sparsity promoter, the cascaded channel and the failure mask can be recovered simultaneously.

\subsection{Performance Evaluation}
In the following, we evaluate the performance of proposed diagnostic techniques.
The results shown are derived from 500 independent Monte-Carlo simulations.

\begin{figure}
	\centering
	\includegraphics[scale=0.4]{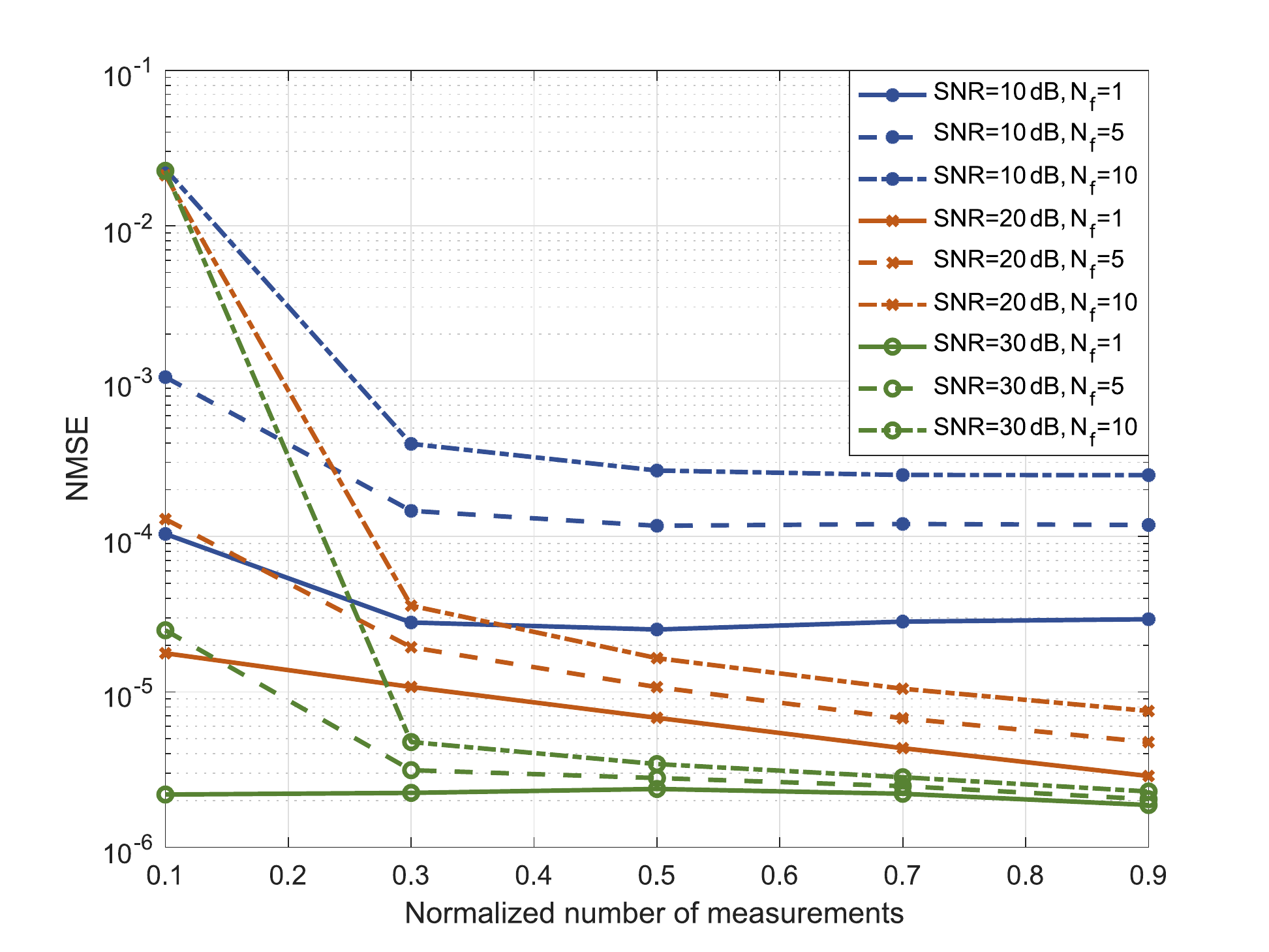}
	\caption{NMSE performance of the diagnosis with full CSI, where $N_\mathrm{f}$ is the number of faulty reflecting elements.}
	\label{F1}
\end{figure}

We first evaluate the performance of diagnosis with full CSI.
The NMSE versus the normalized number of measurements under different numbers of faulty reflecting elements and SNRs is shown in Fig. \ref{F1}.
It can be observed that about $K=0.3N$ measurements are sufficient for the diagnosis with full CSI.
Besides, a higher SNR and/or a lower number of faulty reflecting elements will result in better performance.
We can also observe that the number of faulty reflecting elements has a major impact on the performance when the number of measurements is extremely low ($K=0.1N$), while the SNR dominates the lower bound of the NMSE when the measurement number becomes sufficiently large.

\begin{figure}
	\centering
	\includegraphics[scale=0.4]{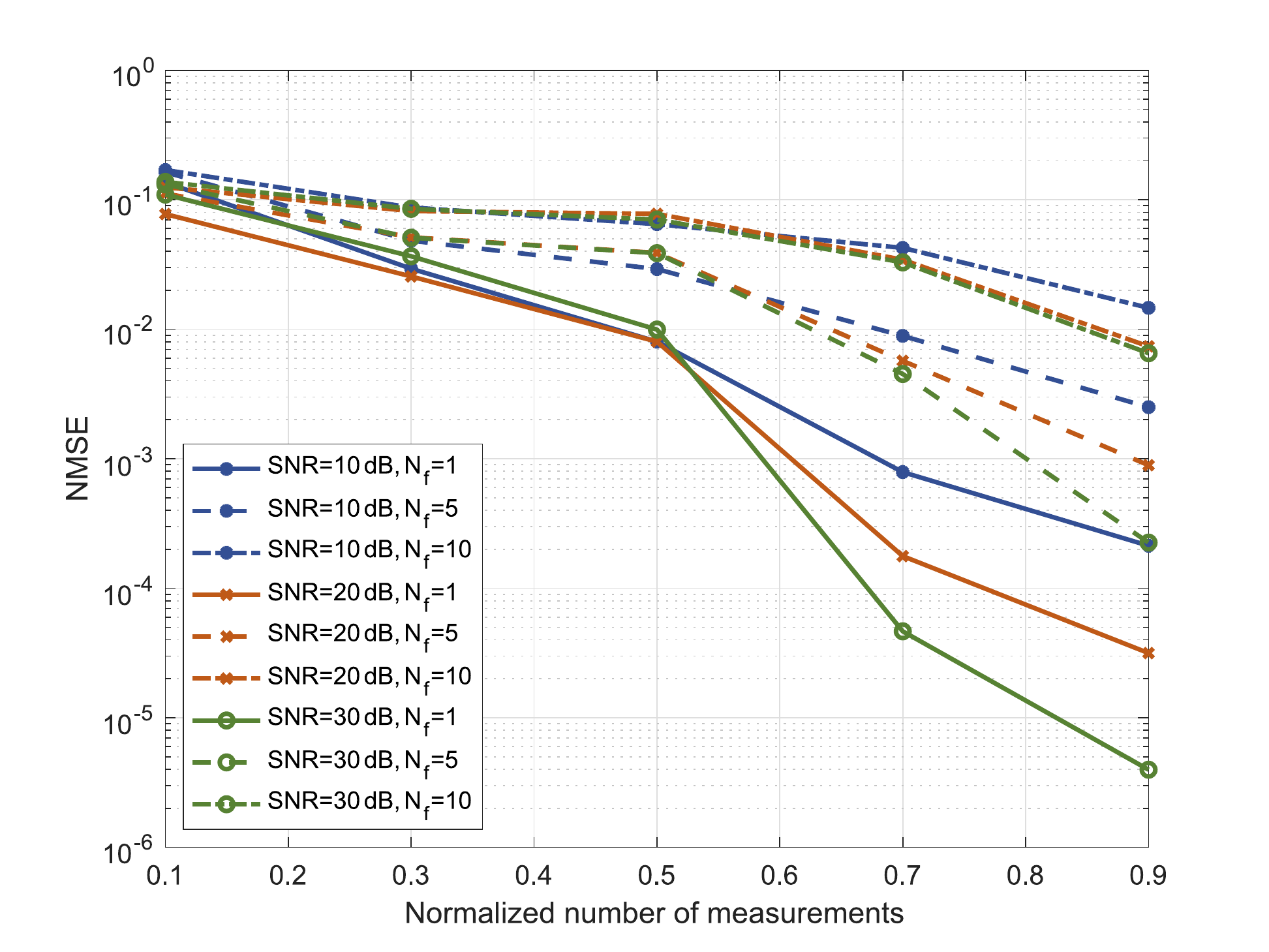}
	\caption{NMSE performance of the diagnosis with partial CSI, where $N_\mathrm{f}$ is the number of faulty reflecting elements.}
	\label{F2}
\end{figure}

Next, we evaluate the performance of diagnosis with partial CSI.
Setting the number of sub-paths in the TX-IRS channel to $L_\mathrm{TX} = 4$, the NMSE versus the normalized number of measurements under different numbers of faulty reflecting elements and SNRs is shown in Fig. \ref{F2}.
Comparing with the diagnosis with full CSI, the diagnosis with partial CSI requires more measurements to deliver satisfactory performance.
In addition, the impact of SNR appears when the number of measurements is sufficiently large, while the performance is sensitive to the number of faulty reflecting elements under all numbers of measurements.

\begin{figure}
	\centering
	\includegraphics[scale=0.4]{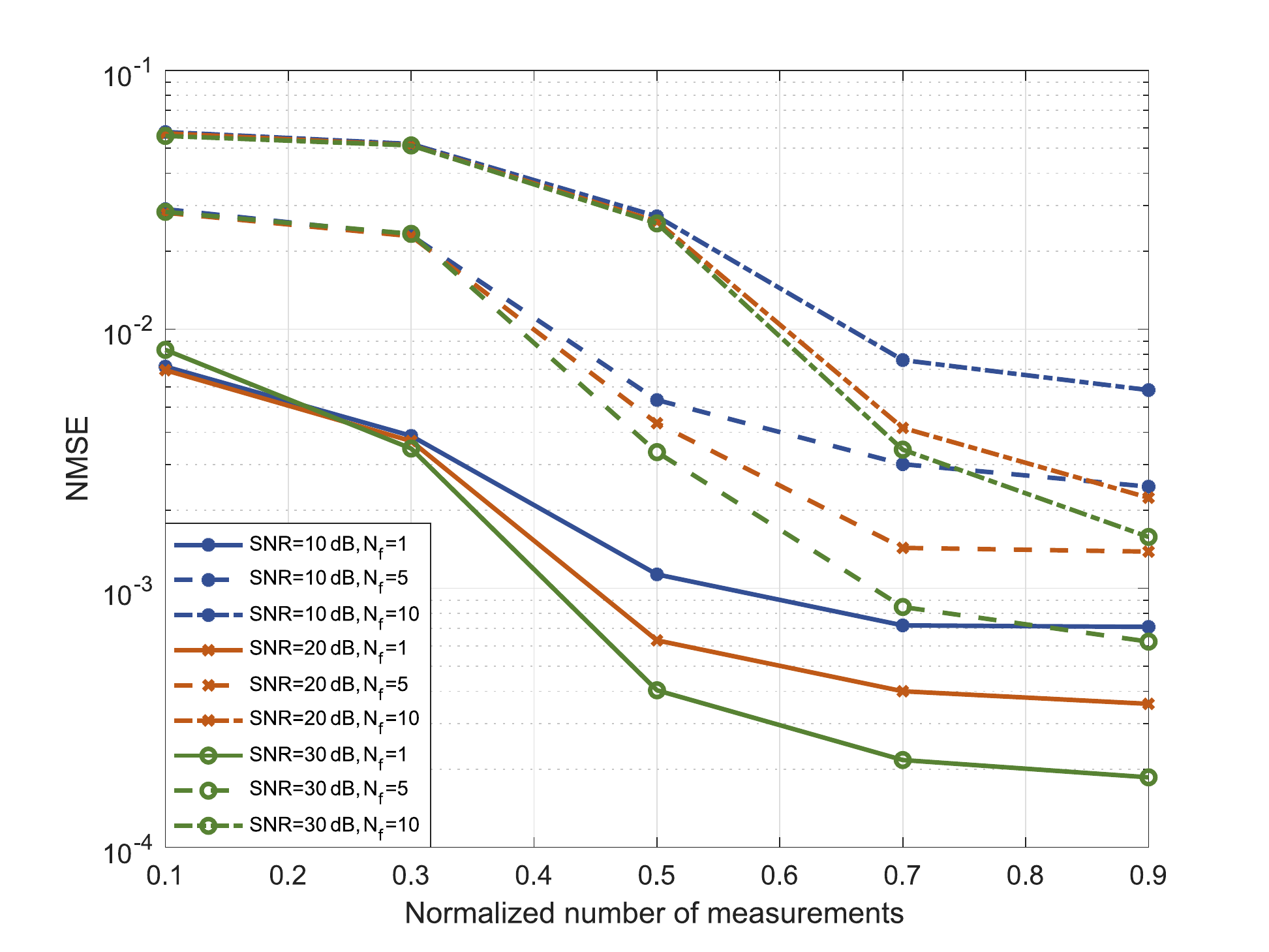}
	\caption{NMSE performance of the diagnosis without CSI, where $N_\mathrm{f}$ is the number of faulty reflecting elements.}
	\label{F3}
\end{figure}

In the following, we evaluate the performance of diagnosis without CSI.
Setting the number of sub-paths to $L_\mathrm{TX} = L_\mathrm{RX} = 4$, we plot the NMSE versus the normalized number of measurements under different numbers of faulty reflecting elements and SNRs in Fig. \ref{F3}.
Similar to the previous results, the impact of SNR becomes significant when the measurements are sufficient, while the number of faulty reflecting elements has a major impact across all numbers of measurements.
Besides, the diagnosis without CSI has better performance in the regime of low measurement number compared with the diagnosis with partial CSI.
The reason may be that the ANM-based diagnosis can tolerate a higher channel sparsity than the cSLRMR-based diagnosis, which reduces the requirement on the measurement number.

\begin{figure}
	\centering
	\includegraphics[scale=0.4]{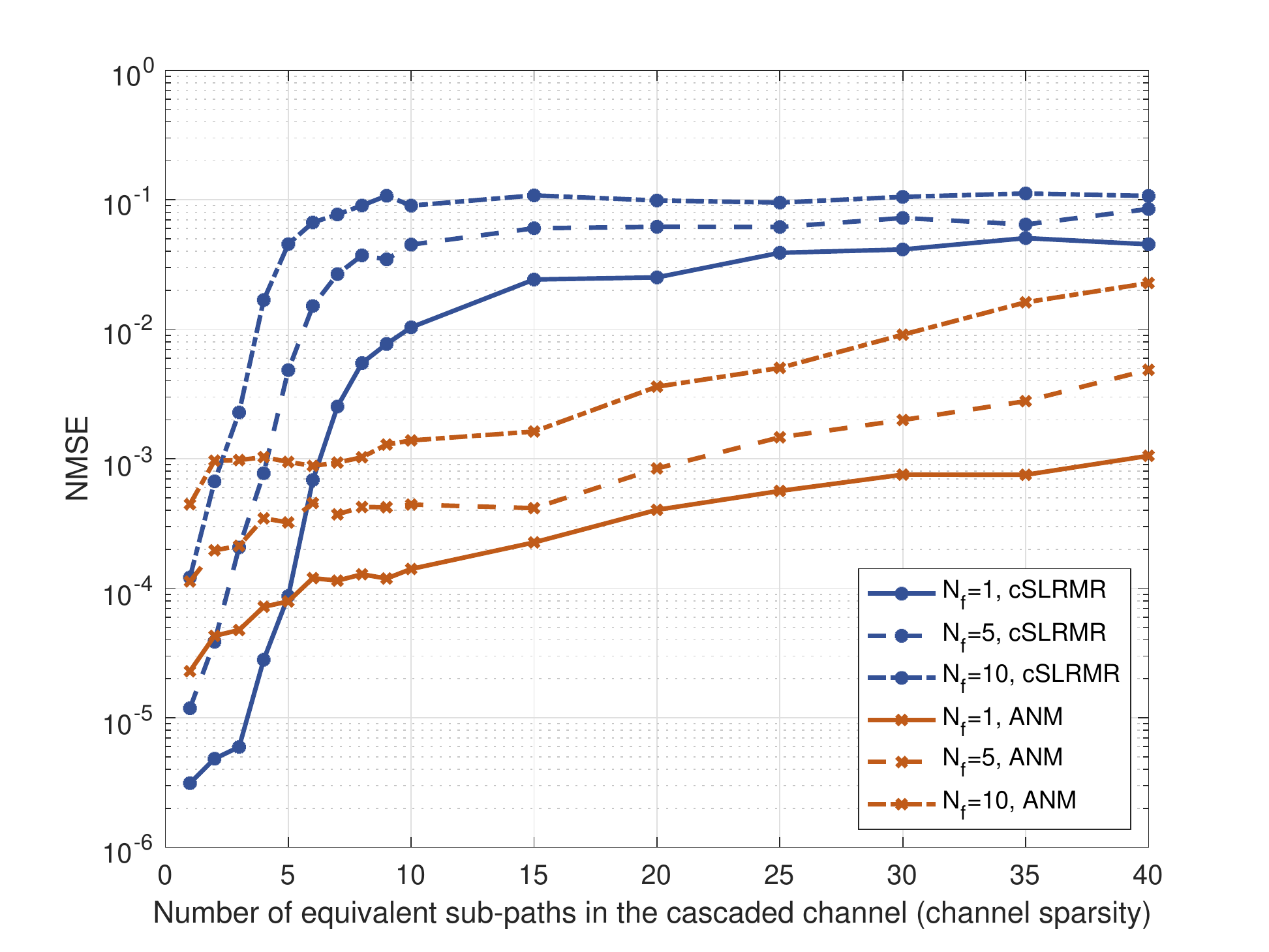}
	\caption{Comparison of the NMSE performance between cSLRMR-based diagnosis and ANM-based diagnosis, where $N_\mathrm{f}$ is the number of faulty reflecting elements.}
	\label{F4}
\end{figure}

To illustrate the tolerance against multipath scattering of the ANM-based diagnostic technique (diagnosis without CSI), Fig. \ref{F4} compares the NMSE performance of the ANM-based diagnostic technique and the diagnosis technique using the cSLRMR under different numbers of channel sub-paths and faulty reflecting elements, where $K=0.8N$, $\text{SNR}=30$ dB, and all channels are assumed unknown.
It can be observed that the diagnostic technique using the cSLRMR applies to the case without CSI provided that the cascaded channel is sufficiently sparse.
However, as the number of sub-paths increases, the performance of cSLRMR-based diagnostic technique degrades rapidly.
In contrast, the ANM-based diagnostic technique shows strong robustness against the multipath scattering and performs well even when the channel matrix is full-rank ($L\ge\min(H,W)$).
Since the failure sparsity increases as the number of faulty reflecting elements grows, all diagnostic techniques achieve better NMSE performance when the fault proportion is low.

\begin{figure}
	\centering
	\includegraphics[scale=0.4]{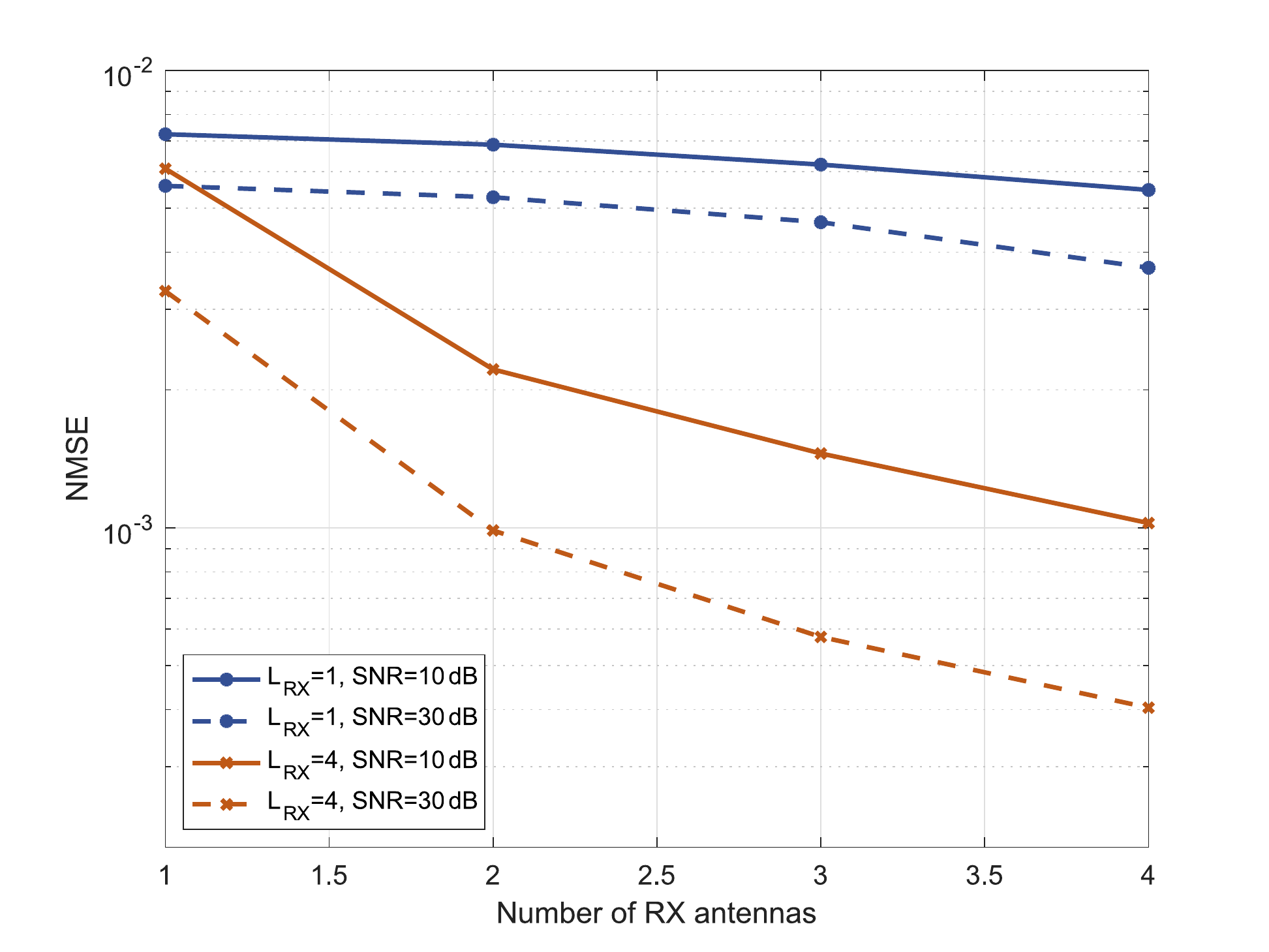}
	\caption{NMSE performance of the ANM-based diagnostic technique using multiple antennas at the RX.}
	\label{F5}
\end{figure}

In Fig. \ref{F5}, we evaluate the effect of multiple antennas at the RX under different numbers of sub-paths in the IRS-RX channel, where $K=0.5N  $, $N_\mathrm{f}=5$, and the number of equivalent sub-paths in the cascaded channel is set to $L = L_\mathrm{TX} L_\mathrm{RX} = 16$ to maintain a constant channel sparsity.
It can be observed that multiple antennas at the RX significantly improve the performance when there is sufficient multipath scattering in the IRS-RX channel ($L_\mathrm{RX} > 1$).
The comparison under different SNRs also verifies that the effect of multiple antennas is mainly the diversity gain rather than the SNR gain.

\begin{figure}
	\centering
	\includegraphics[scale=0.4]{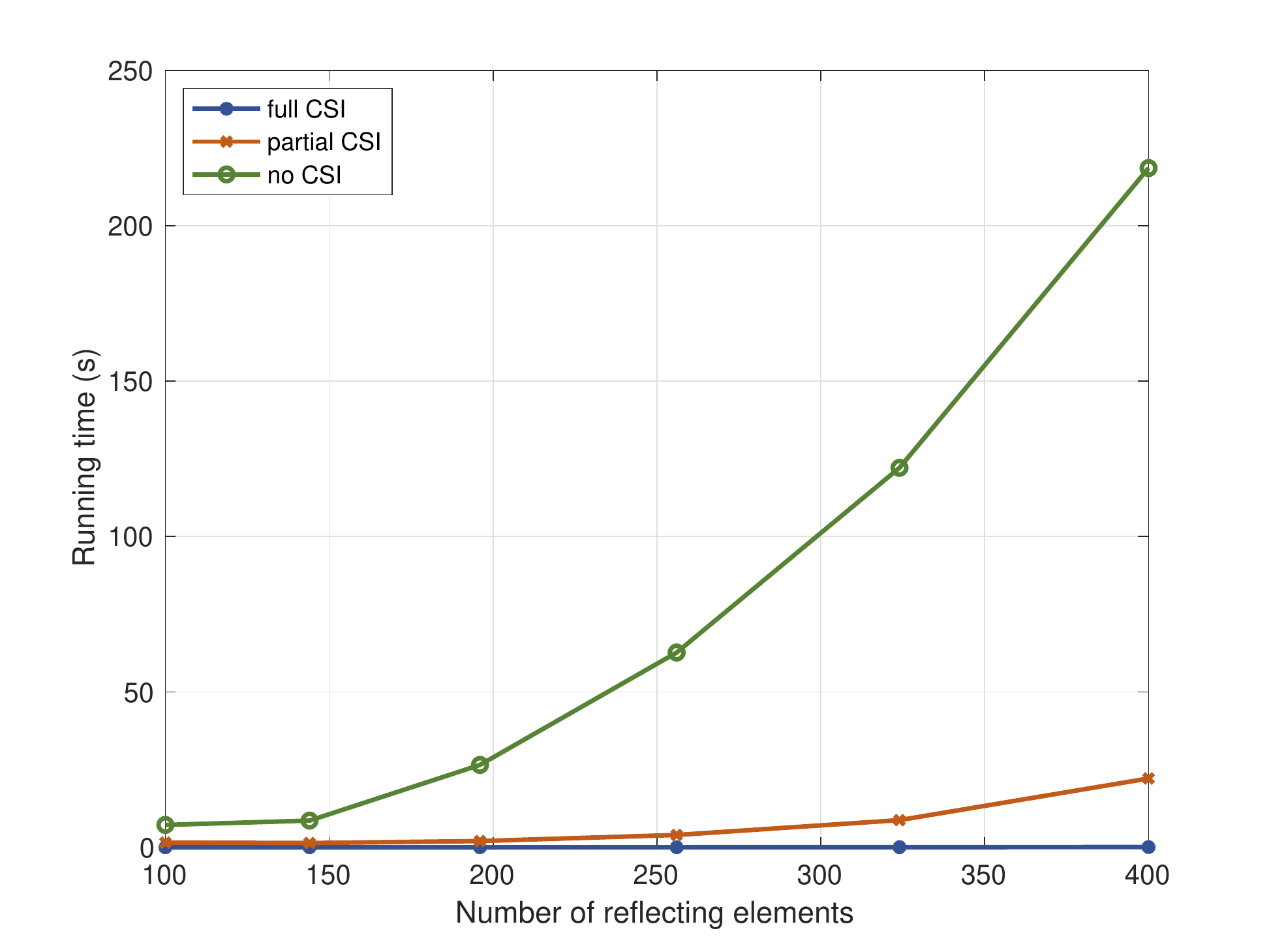}
	\caption{Comparison of the running time between different diagnostic techniques.}
	\label{F6}
\end{figure}

Finally, we evaluate the time complexity of different diagnostic techniques.
Fig. \ref{F6} plots the running time versus the number of reflecting elements $N$, where all MATLAB codes are executed on a desktop PC with an Intel Core i7-4790K CPU running at 4.0 GHz.
The results show that there is a considerable gap in running time between different diagnostic techniques, just as they have different requirements of the CSI.
Although the ANM-based diagnostic technique (diagnosis without CSI) also applies to the case of full/partial CSI, we recommend selecting a proper diagnostic technique based on the CSI availability to avoid unnecessary computation.

\section{Conclusions}
\label{sec:conclu}
In this paper, we have developed several diagnostic techniques for IRS system.
We have shown that the CSI availability is the most challenging factor for diagnosis since the channel and the failure are intimately coupled.
To perform diagnosis in different scenarios, three diagnostic techniques have been developed, which have different requirements for CSI as well as different computational complexities.
In the first case where full CSI is available, the impact of the channel can be eliminated by subtracting the reference signal from the received signal, and the diagnosis problem has been formulated as a compressed sensing problem by exploiting the failure sparsity.
In the second case where only partial CSI is available, the channel-failure decoupling has been achieved by jointly exploiting the low-rank property of the mmWave channel and the sparsity of the failure.
The third case where no CSI is available is the most challenging since the cascaded channel may not be sparse in conventional dictionaries, which is a unique challenge introduced by IRS hardware characteristics.
To cope with this issue, we have shown that the cascaded channel can be sparsely represented in the continuous 2D Fourier dictionary, and introduced the atomic norm to induce the sparsity of cascaded channel.
Then, an ANM-based diagnostic technique has been proposed to decouple the cascaded channel and the failure.
Two ADMM-based algorithms have also been proposed to accelerate the computation.
Numerical simulations have demonstrated that the proposed diagnostic techniques can deliver satisfactory performance.

\appendices

\section{An Efficient Algorithm for Solving \eqref{ANMfinal}}
\label{sec:ADMM}
We first rewrite \eqref{ANMfinal} into an ADMM-compliant form:
\begin{equation}
\begin{aligned}
\{ \hat{\mathbf{h}}, \hat{\mathbf{d}}\}
=& \arg \min_{\mathbf{h, d, u}, v} 
\frac{1}{2}\Vert \mathbf{y-F(h + d)}\Vert_2^2 \\
&+ \frac{\tau_1}{2} \left( \frac{1}{N} \mathrm{tr}\big(T(\mathbf{u})\big) + v \right)
+ \lambda_3 \Vert \mathbf{d} \Vert_1, \\
&\mathrm{s.t.} \hspace{0.5em}
\mathbf{Z} = 
\begin{bmatrix}
T(\mathbf{u}) & \mathbf{h} \\
\mathbf{h}^\mathrm{H} & v
\end{bmatrix}
\succeq 0,
\end{aligned}
\label{ADMM1}
\end{equation}
where $\mathbf{Z} \in \mathbb{C}^{(N+1) \times (N+1)}$ is an auxiliary matrix.

The augmented Lagrangian function for \eqref{ADMM1} is of the form
\begin{equation}
\label{ADMM2}
\begin{aligned}
&\mathcal{L}_\rho(v, \mathbf{u, h, d, Z}, \boldsymbol{\Lambda}) = \frac{1}{2}\Vert \mathbf{y-F(h + d)}\Vert_2^2 \\
&+ \frac{\tau_1}{2} \left( \frac{1}{N} \mathrm{tr}\big(T(\mathbf{u})\big) + v \right)
+ \lambda_3 \Vert \mathbf{d} \Vert_1 \\
&+ \left\langle \boldsymbol{\Lambda}, \mathbf{Z} - \begin{bmatrix}
T(\mathbf{u}) & \mathbf{h} \\
\mathbf{h}^\mathrm{H} & v
\end{bmatrix} \right \rangle
+ \frac{\rho}{2} \left\Vert \mathbf{Z} - \begin{bmatrix}
T(\mathbf{u}) & \mathbf{h} \\
\mathbf{h}^\mathrm{H} & v
\end{bmatrix} \right\Vert_\mathrm{F}^2,
\end{aligned}
\end{equation}
where $\boldsymbol{\Lambda} \in \mathbb{C}^{(N+1) \times (N+1)}$ is the Lagrangian multiplier, $\rho$ is a penalty parameter, and $\langle \cdot, \cdot \rangle$ denotes the real inner product.

The ADMM algorithm minimizes the augmented Lagrangian function by iteratively performing the following updates:
\begin{equation}
\begin{aligned}
	v^{(l+1)} &= \arg \min_{v} \mathcal{L}_\rho(v, \mathbf{u}^{(l)}, \mathbf{h}^{(l)}, \mathbf{d}^{(l)}, \mathbf{Z}^{(l)}, \boldsymbol{\Lambda}^{(l)}), \\
	\mathbf{u}^{(l+1)} &= \arg \min_\mathbf{u} \mathcal{L}_\rho(v^{(l+1)}, \mathbf{u}, \mathbf{h}^{(l)}, \mathbf{d}^{(l)}, \mathbf{Z}^{(l)}, \boldsymbol{\Lambda}^{(l)}), \\
	\mathbf{h}^{(l+1)} &= \arg \min_\mathbf{h} \mathcal{L}_\rho(v^{(l+1)}, \mathbf{u}^{(l+1)}, \mathbf{h}, \mathbf{d}^{(l)}, \mathbf{Z}^{(l)}, \boldsymbol{\Lambda}^{(l)}), \\
	\mathbf{d}^{(l+1)} &= \arg \min_\mathbf{d} \mathcal{L}_\rho(v^{(l+1)}, \mathbf{u}^{(l+1)}, \mathbf{h}^{(l+1)}, \mathbf{d}, \mathbf{Z}^{(l)}, \boldsymbol{\Lambda}^{(l)}), \\
	\mathbf{Z}^{(l+1)} &= \arg \min_\mathbf{Z} \mathcal{L}_\rho(v^{(l+1)}, \mathbf{u}^{(l+1)}, \mathbf{h}^{(l+1)}, \mathbf{d}^{(l+1)}, \mathbf{Z}, \boldsymbol{\Lambda}^{(l)}), \\
	\boldsymbol{\Lambda}^{(l+1)} &= \boldsymbol{\Lambda}^{(l)} + \rho \left( \mathbf{Z}^{(l+1)} - \begin{bmatrix}
		T\left(\mathbf{u}^{(l+1)}\right) & \mathbf{h}^{(l+1)} \\
		{\mathbf{h}^{(l+1)}}^\mathrm{H} & v^{(l+1)}
	\end{bmatrix} \right),
\end{aligned}
\end{equation}
where the superscript $(l)$ denotes the $l$-th iteration.

To perform the above updates in an efficient way, we first introduce the partitions
\begin{equation}
\mathbf{Z}^{(l)} =
\begin{bmatrix}
	\mathbf{Z}^{(l)}_0 & \mathbf{z}^{(l)}_1 \\
	{\mathbf{z}^{(l)}_1}^\mathrm{H} & Z^{(l)}_{N+1, N+1}
\end{bmatrix},
\boldsymbol{\Lambda}^{(l)} =
\begin{bmatrix}
	\boldsymbol{\Lambda}^{(l)}_0 & \boldsymbol{\lambda}^{(l)}_1 \\
	{\boldsymbol{\lambda}^{(l)}_1}^\mathrm{H} & \Lambda^{(l)}_{N+1, N+1}
\end{bmatrix}.
\end{equation}

The augmented Lagrangian function is convex and differentiable with respect to the variables $v$, $\mathbf{u}$, and $\mathbf{h}$.
Setting their gradients to zero, these updates admit closed-form expressions as
\begin{equation}
\label{update_u}
\begin{aligned}
v^{(l+1)} &= Z^{(l)}_{N+1, N+1} + \frac{1}{\rho} \left(\Lambda^{(l)}_{N+1, N+1} - \frac{\tau_1}{2} \right), \\
\mathbf{u}^{(l+1)} &= \boldsymbol{\Psi}^{-1} \left( T^*\left( \mathbf{Z}^{(l)}_0 + \frac{1}{\rho} \boldsymbol{\Lambda}^{(l)}_0 \right) \right) - \frac{\tau_1}{2 \rho} \mathbf{e}_1, \\
\mathbf{h}^{(l+1)} &= (\mathbf{F}^\mathrm{H} \mathbf{F} + 2\rho \mathbf{I}_{N})^{-1} \left( \mathbf{F}^\mathrm{H}(\mathbf{y} - \mathbf{F} \mathbf{d}^{(l)}) + 2\boldsymbol{\lambda}^{(l)}_1 + 2 \rho \mathbf{z}^{(l)}_1 \right),
\end{aligned}
\end{equation}
where
\begin{equation}
\begin{aligned}
\boldsymbol{\Psi} =& \mathrm{diag}([HW, H(W-1), \cdots, H, [H-1, H-2, \cdots, 1] \\
& \otimes [1, 2, \cdots, W, W-1, W-2, \cdots, 1]]) \in \mathbb{R}^{N_u \times N_u},
\end{aligned}
\end{equation}
$\mathbf{e}_1$ is a zero vector with the first entry being one, and $T^*(\cdot)$ is the adjoint of $T(\cdot)$.
For a given matrix $\mathbf{Q} \in \mathbb{C}^{N \times N}$,
\begin{equation}
T^*(\mathbf{Q}) = [\mathrm{tr}((\boldsymbol{\Theta}_{k_1} \otimes \boldsymbol{\Theta}_{k_2})\mathbf{Q}) | (k_1,k_2) \in \mathsf{H} ]^\mathrm{T},
\end{equation}
where $\boldsymbol{\Theta}_k$ is an elementary Toeplitz matrix with ones on the $k$-th diagonal and zeros elsewhere, $\mathsf{H}$ is a halfspace of $(H-1, W-1)$, and $(k_1,k_2) \in \mathsf{H}$ means $k_2 \in \{0, 1, \cdots, W-1\}$ when $k_1=0$ and $k_2 \in \{-(W-1), -(W-2), \cdots, W-1\}$ when $k_1 \in \{1, 2, \cdots, H-1\}$.

The update of $\mathbf{d}$ can be written as
\begin{equation}
\begin{aligned}
\mathbf{d}^{(l+1)} &= \arg \min_\mathbf{d} \mathcal{L}_\rho(v^{(l+1)}, \mathbf{u}^{(l+1)}, \mathbf{h}^{(l+1)}, \mathbf{d}, \mathbf{Z}^{(l)}, \boldsymbol{\Lambda}^{(l)}) \\
&= \arg \min_\mathbf{d} \frac{1}{2}\Vert \mathbf{y} - \mathbf{F}(\mathbf{h}^{(l+1)} + \mathbf{d}) \Vert_2^2 + \lambda_3 \Vert \mathbf{d} \Vert_1 \\
&= \arg \min_\mathbf{d} \frac{1}{2}\Vert (\mathbf{y} - \mathbf{F} \mathbf{h}^{(l+1)}) - \mathbf{F} \mathbf{d} \Vert_2^2 + \lambda_3 \Vert \mathbf{d} \Vert_1,
\end{aligned}
\end{equation}
which is the well-known LASSO problem and can be solved efficiently by the algorithm proposed in \cite{boyd2011distributed}.

Finally, the update of $\mathbf{Z}$
\begin{equation}
\begin{aligned}
\mathbf{Z}^{(l+1)} &= \arg \min_\mathbf{Z} \mathcal{L}_\rho(v^{(l+1)}, \mathbf{u}^{(l+1)}, \mathbf{h}^{(l+1)}, \mathbf{d}^{(l+1)}, \mathbf{Z}, \boldsymbol{\Lambda}^{(l)}) \\
&= \arg \min_{\mathbf{Z} \succeq 0} \left\Vert \mathbf{Z} - 
\left(
\underbrace{
\begin{bmatrix}
T\left(\mathbf{u}^{(l+1)}\right) & \mathbf{h}^{(l+1)} \\
{\mathbf{h}^{(l+1)}}^\mathrm{H} & v^{(l+1)}
\end{bmatrix}
- \frac{1}{\rho}\boldsymbol{\Lambda}^{(l)}
}_{=\mathbf{G}}
\right)
\right\Vert_\mathrm{F}^2
\end{aligned}
\end{equation}
amounts to projecting the Hermitian matrix $\mathbf{G}$ onto the positive semidefinite cone, which can be performed by computing the eigenvalue decomposition of $\mathbf{G}$ and setting all negative eigenvalues to zero.

We now analyze the time complexity of the above algorithm.
The computational complexity is mainly contributed by the updates of $\mathbf{u, h, d}$, and $\mathbf{Z}$.
Note that the update of $\mathbf{u}$ can be simplified by avoiding unnecessary matrix multiplications.
In \eqref{update_u}, the calculation of $\boldsymbol{\Psi}^{-1}$ is independent of the input and thus can be cached before the iteration.
The adjoint operator $T^*(\cdot)$ is essentially an ``extract and sum" operator, where each output element of $T^*(\cdot)$ is the sum of multiple specific elements of the input, whose indices can also be cached before the iteration.
Therefore, the time complexity of the update of $\mathbf{u}$ can be reduced to $O(N_u)$ (the multiplication of an $N_u \times N_u$ diagonal matrix $\boldsymbol{\Psi}^{-1}$ and an $N_u$-dimensional vector $T^*(\cdot)$).
The update of $\mathbf{h}$ has time complexity of $O(N^3)$ due to the matrix inversion $(\mathbf{F}^\mathrm{H} \mathbf{F} + 2\rho \mathbf{I}_{N})^{-1}$.
The update of $\mathbf{d}$ (the LASSO problem) also has time complexity of $O(N^3)$ introduced by $N$-dimensional matrix inversion \cite{boyd2011distributed}.
The update of $\mathbf{Z}$ amounts to eigenvalue decomposition, whose time complexity is upper bounded by $O(N^3)$.
Combining the above results, the overall complexity of the proposed algorithm is $O(N^3)$.

\section{An Efficient Algorithm for Solving \eqref{ANMfinalMMV}}
\label{sec:ADMM2}

An efficient ADMM-based algorithm for solving \eqref{ANMfinalMMV} is presented in Algorithm \ref{algoMMV}.
One can refer to Appendix \ref{sec:ADMM} for a similar derivation.

Under the MMV model, the time complexity of the proposed algorithm for solving \eqref{ANMfinalMMV} increases to $O\left((N + N_\mathrm{RX})^3\right)$ due to the increased size of the eigenvalue decomposition problem (the update of $\mathbf{Z}$).

\begin{algorithm}[h]
	\caption{An efficient ADMM-based algorithm for solving \eqref{ANMfinalMMV}}
	\label{algoMMV}
	Initialization: $\mathbf{Z}^{(0)} = \mathbf{0}$, $\boldsymbol{\Lambda}^{(0)} = \mathbf{0}$, $\mathbf{D}^{(0)} = \mathbf{0}$; \\
	\While{not converged}{
		\nl $\mathbf{V}^{(l+1)} = \mathbf{Z}^{(l)}_{N_\mathrm{RX}} + \frac{1}{\rho} \left(\boldsymbol{\Lambda}^{(l)}_{N_\mathrm{RX}} - \frac{\tau_2}{2} \mathbf{I}_{N_\mathrm{RX}} \right)$; \\
		
		\nl $\mathbf{u}^{(l+1)} = \boldsymbol{\Psi}^{-1} \left( T^*\left( \mathbf{Z}^{(l)}_0 + \frac{1}{\rho} \boldsymbol{\Lambda}^{(l)}_0 \right) \right) - \frac{\tau_2}{2 \rho} \mathbf{e}_1$; \\
		
		\nl $\mathbf{H}^{(l+1)} = (\mathbf{F}^\mathrm{H} \mathbf{F} + 2\rho \mathbf{I}_{N})^{-1} \left( \mathbf{F}^\mathrm{H}(\mathbf{Y} - \mathbf{F} \mathbf{D}^{(l)}) + 2\boldsymbol{\Lambda}^{(l)}_1 + 2 \rho \mathbf{Z}^{(l)}_1 \right)$; \\
		
		\nl $\mathbf{D}^{(l+1)} = \arg \min_\mathbf{D} \frac{1}{2} \Vert \left( \mathbf{Y} - \mathbf{F} \mathbf{H}^{(l)} \right) - \mathbf{FD} \Vert_2^2 + \lambda_4 \Vert \mathbf{D} \Vert_{2,1}$, solved by the M-FOCUSS algorithm \cite{cotter2005sparse}; \\
		
		\nl Perform eigenvalue decomposition on
		$
		\begin{bmatrix}
			T\left(\mathbf{u}^{(l+1)}\right) & \mathbf{H}^{(l+1)} \\
			{\mathbf{H}^{(l+1)}}^\mathrm{H} & \mathbf{V}^{(l+1)}
		\end{bmatrix}
		- \frac{1}{\rho}\boldsymbol{\Lambda}^{(l)}
		$
		and set all negative eigenvalues to zero, yielding $\mathbf{Z}^{(l+1)}$; \\
		
		\nl $\boldsymbol{\Lambda}^{(l+1)} = \boldsymbol{\Lambda}^{(l)} + \rho \left( \mathbf{Z}^{(l+1)} - 
		\begin{bmatrix}
			T\left(\mathbf{u}^{(l+1)}\right) & \mathbf{H}^{(l+1)} \\
			{\mathbf{H}^{(l+1)}}^\mathrm{H} & \mathbf{V}^{(l+1)}
		\end{bmatrix}
		\right)$;
	}
	
	where
	
	$
	\mathbf{Z}^{(l)} = 
	\begin{bmatrix}
		\mathbf{Z}^{(l)}_{0} \in \mathbb{C}^{N \times N} & \mathbf{Z}^{(l)}_{1} \in \mathbb{C}^{N \times N_\mathrm{RX}} \\
		{\mathbf{Z}^{(l)}_{1}}^\mathrm{H} \in \mathbb{C}^{N_\mathrm{RX} \times N} & \mathbf{Z}^{(l)}_{N_\mathrm{RX}} \in \mathbb{C}^{N_\mathrm{RX} \times N_\mathrm{RX}}
	\end{bmatrix}
	$,
	$
	\boldsymbol{\Lambda}^{(l)} = 
	\begin{bmatrix}
		\boldsymbol{\Lambda}^{(l)}_{0} \in \mathbb{C}^{N \times N} & \boldsymbol{\Lambda}^{(l)}_{1} \in \mathbb{C}^{N \times N_\mathrm{RX}} \\
		{\boldsymbol{\Lambda}^{(l)}_{1}}^\mathrm{H} \in \mathbb{C}^{N_\mathrm{RX} \times N} & \boldsymbol{\Lambda}^{(l)}_{N_\mathrm{RX}} \in \mathbb{C}^{N_\mathrm{RX} \times N_\mathrm{RX}}
	\end{bmatrix}
	$.
\end{algorithm}

\bibliographystyle{IEEEtran}
\bibliography{ref}

\end{document}